\documentclass{article}
\usepackage[a4paper, top=4cm, bottom=3cm, inner=3cm, outer=3cm]{geometry}

\usepackage{fancyhdr}

\usepackage[utf8]{inputenc}
\usepackage{lmodern}
\usepackage[T1]{fontenc}

\usepackage{amsmath, amssymb, amsthm, amsfonts}

\usepackage[english]{babel}

\usepackage{bbm}
\usepackage[retainorgcmds]{IEEEtrantools}
\usepackage{thm-restate}
\usepackage{enumerate}

\usepackage{tikz}
\usetikzlibrary{calc, arrows, decorations.markings, positioning, fit,
  shapes.geometric, arrows.meta}
\tikzset{
  >=stealth'
}
\tikzstyle{vecArrow} = [thick, decoration={markings,mark=at position
  1 with {\arrow[semithick]{open triangle 60}}},
  double distance=1.4pt, shorten >= 5.5pt,
  preaction = {decorate},
  postaction = {draw,line width=1.4pt, white,shorten >= 4.5pt}]
\tikzstyle{myArrow} = [thick, ->]

\usepackage{listings}
\usepackage{verbatim}

\theoremstyle{plain}
        \newtheorem{theorem}{Theorem}
        
        \newtheorem{corollary}[theorem]{Corollary}
        \newtheorem{claim}[theorem]{Claim}

\theoremstyle{definition}
        \newtheorem{definition}[theorem]{Definition}
        \newtheorem{example}[theorem]{Example}

\theoremstyle{remark}
        \newtheorem{remark}[theorem]{Remark}

\DeclareMathOperator*{\EE}{E}

\DeclareMathOperator{\diam}{diam}

\DeclareMathOperator{\3SAT}{3-SAT}
\DeclareMathOperator{\SAT}{SAT}
\DeclareMathOperator{\2SAT}{2-SAT}
\DeclareMathOperator{\TQBF}{TQBF}
\DeclareMathOperator{\BAY}{BINARY-ACTION}

\DeclareMathOperator{\BAYAB}{BINARY-BELIEF}

\newcommand{\NP}{\mathsf{NP}}

\newcommand{\sT}{\mathsf{T}}
\newcommand{\sF}{\mathsf{F}}
\newcommand{\OBS}{\mathsf{OBS}}
\newcommand{\EPS}{\mathsf{EPS}}
\newcommand{\EVAL}{\mathsf{EVAL}}
\newcommand{\PSPACE}{\mathsf{PSPACE}}
\newcommand{\HASHP}{\mathsf{\#P}}
\newcommand{\eps}{\varepsilon}

\definecolor{DSgray}{cmyk}{0,0,0,0.7}
\definecolor{DSred}{cmyk}{0,0.7,0,0.7}

\usepackage[pdftex]{hyperref}

\title{Reasoning in Bayesian Opinion Exchange Networks Is PSPACE-Hard}

\author{
  Jan Hązła\thanks{Massachusetts Institute of Technology, {\tt jhazla@mit.edu}.
  } \and
  Ali Jadbabaie\thanks{MIT, {\tt jadbabai@mit.edu}.} \and
  Elchanan Mossel\thanks{MIT, {\tt elmos@mit.edu}.
    E.M.~and J.H.~are partially supported by awards ONR N00014-16-1-2227,  
    NSF CCF-1665252 and DMS-1737944. } \and
  M. Amin Rahimian\thanks{MIT, {\tt rahimian@mit.edu}.}
}

\begin{document}

\maketitle

\begin{abstract}
  We study the Bayesian model of opinion exchange of fully rational
  agents arranged on a network.
  In this model, the agents receive private signals that are indicative of
  an unkown state of the world. 
  Then, they repeatedly announce the state of the world
  they consider most likely to their neighbors, at the same time updating
  their beliefs based on their neighbors' announcements.
 
  This model is extensively studied in economics since the work of Aumann
  (1976)
  and Geanakoplos and Polemarchakis
  (1982).
  It is known that the agents eventually agree with high probability on any
  network.
  It is often argued that the computations needed by agents in this model
  are difficult, but prior to our results there was no rigorous work showing
  this hardness. 

  We show that it is  $\PSPACE$-hard for the agents to compute
  their actions in this model. Furthermore, we show that it is equally
  difficult even to approximate an agent's posterior: It is $\PSPACE$-hard
  to distinguish between the posterior being almost entirely concentrated
  on one state of the world or another. 
\end{abstract}

\nocite{Aum76, GP82}

\section{Introduction}
\label{sec:intro}

\paragraph{Background}
The problem of dynamic opinion exchange is an important field of study
in economics, with its roots reaching as far as the Condorcet's jury theorem
and, in the Bayesian context, Aumann's agreement theorem.
Economists use different opinion exchange models as inspiration for
explaining interactions and decisions of market participants. More generally,
there is extensive interest in studying how social agents exchange information,
form opinions and use them as a basis to make decisions.
For a more comprehensive introduction to the subject we refer 
to surveys addressed to economists \cite{AO11} and mathematicians \cite{MT17}.

Many models have been proposed and researched, with the properties studied
including, among others,
if the agents converge to the same opinion, the rate of such convergence,
and if the consensus decision is optimal with high probability
(this is called \emph{learning}). Two interesting aspects of the differences
between models are rules for updating agents' opinions (e.g., fully rational
or heuristic) and presence of network structure.

For example, in settings where the updates are assumed to be rational
(Bayesian), there is extensive study of models where the agents act in
sequence
(see, e.g., \cite{Ban92, BHW92, SS00, ADLO11} for a non-exhaustive
selection of works that consider phenomena of
\emph{herding} and \emph{information cascades}),
as well as models with agents arranged in a network
and repeatedly exchanging opinions as time progresses
(see some references below). In this work we are interested in the
network models, arguably becoming more and more relevant given the ubiquity of
networks in modern society.

On the other hand, similar questions are studied for models with
so-called \emph{bounded rationality}, where the Bayesian updates are replaced
with simpler, heuristic rules. Some well-known examples include DeGroot model
\cite{DeG74, GJ10}, the voter model \cite{CS73, HL75} and other
related variants \cite{BG98, AOP10}.

One commonly accepted reason for studying bounded rationality is that,
especially in the network case, Bayesian updates
become so complicated as to make fully rational behavior intractable,
and therefore unrealistic. However, we are not aware of previous theoretical
evidence or formalization of that assertion. Together with another paper
of the same authors addressed to economists \cite{JMR17}, we consider this
work as a development in that direction.

More precisely, we show that computing an agent's opinion
in one of the most important and studied Bayesian network models is
$\PSPACE$-hard. 
Furthermore, it is $\PSPACE$-hard even to
approximate the rational opinion in any meaningful way.
This improves on our $\NP$-hardness result
for the same problem shown in \cite{JMR17}.

\paragraph{Our model and results}
We are concerned with a certain Bayesian model of opinion
exchange and reaching agreement on a network.
We are going to call it the \emph{(Bayesian) binary action model}.
The idea is that there is a network of honest, fully rational agents
trying to learn a binary piece of information, e.g., will the price of an
asset go up or down, or which political party's policies are more beneficial
to the society. We call this information the \emph{state of the world}.
Initially, each agent receives an independent piece of information
(a \emph{private signal}) that is correlated with the state of the world.
According to the principle that ``actions speak louder than
words'', at every time step the agents reveal to their neighbors which
of the two possible states they consider more likely.
On the other hand, we assume that the agents are honest truth-seekers
and always truthfully reveal their preferred state: According
to economic terminology they act \emph{myopically} rather than
strategically.

More specifically, we assume that the
state of the world is encoded in a random variable $\theta \in \{\sT, \sF\}$
(standing for True and False), distributed according to a uniform prior,
shared by all agents. A set of Bayesian agents
arranged on a directed graph $G = (V, E)$ performs a sequence of actions
at discrete times $t = 0,1,2,\ldots$
Before the process starts, each agent $u$ receives a random private signal
$S(u) \in \{0,1\}$. The collection of random variables
$\{S(u): u \in V\}$ is independent conditioned on $\theta$.
The idea is that $S(u) = 1$ indicates a piece of evidence for $\theta = \sT$
and $S(u) = 0$ is evidence favoring $\theta = \sF$.

At each time $t \ge 0$, the agents simultaneously broadcast \emph{actions}
to their neighbors in $G$. The action $A(u, t) \in \{\sT, \sF\}$ is the
best guess for the state of the world by agent $u$ at time $t$:
Letting $\mu(u, t)$ be the respective
Bayesian posterior probability that $\theta = \sT$, the action
$A(u, t) = \sT$ if and only if $\mu(u, t) > 1/2$. In subsequent steps,
agents update their posteriors based on their neighbors' actions
(we assume that everyone is rational, and that this fact and the description
of the model are common knowledge) and broadcast updated actions. The process continues
indefinitely.

We are interested in computational resources required for the agents to
participate in the process described above.
That is, we consider complexity of computing
the action $A(u, t)$ given the private signal $S(u)$ and
history of observations $\{ A(v, t'): v \in \mathcal{N}(u), t' < t\}$,
where $\mathcal{N}(u)$ denotes the set of neighbors of $u$ in $G$.
Our main result is that it is worst-case $\PSPACE$-hard for an agent to
distinguish between cases where $\mu(u,t) \ge 1-\exp(-\Theta(N))$ and
$\mu(u,t) \le \exp(-\Theta(N))$, where $N$ is a naturally defined size of the
problem. As a consequence, it is $\PSPACE$-hard to compute the action
$A(u, t)$.

Note a \emph{hardness of approximation} aspect of our result:
A priori one can imagine a reduction
where it is difficult to compute the action $A(u, t)$ when the Bayesian
posterior is close to the threshold $\mu(u, t) \approx 1/2$.
However, we demonstrate that it is already hard to distinguish between
situations where the posterior is concentrated on one of the extreme values
$\mu(u, t) \approx 0$ (and therefore almost certainly $\theta=\sF$)
and $\mu(u, t) \approx 1$ (and therefore $\theta = \sT$).

Our hardness results carry over to other models.
In particular, they extend to the case where the signals are
continuous, where the prior state of the world is not uniform etc.
We also note that we may assume that the agents are never tied or
close to tied in their posteriors, see Remark~\ref{rem:ties} for more details.

A good deal is known about the model we are considering.
From a paper by Gale and Kariv~\cite{GK03} (with an error
corrected by Rosenberg, Solan and Vieille~\cite{RSV09}, see also similar
analysis of earlier, related models in
\cite{BV82, TA84}) it follows
that if the network $G$ is strongly connected, then the agents eventually
converge to the same action (or they become indifferent). 
The work of Geanakoplos~\cite{Gea94} 
implies that this agreement is
reached in at most $|V| \cdot 2^{|V|}$ time steps.
Furthermore, Mossel, Sly and Tamuz~\cite{MST14} showed that in large undirected
networks with non-atomic signals, learning occurs: The common agreed action is equal to
the state of the world $\theta$, except with probability that goes to zero
with the number of agents.
A good deal remains open, too. For example, it is not known if the
$|V| \cdot 2^{|V|}$ bound on the agreement speed can be improved. 
In this context is also interesting to note the results
of~\cite{MOT16} who consider a special model 
with Gaussian structure and revealed beliefs.
In contrast to the results presented here,
it is shown that in this case,
agents' computations are efficient (polynomial time) and 
convergence time is $O(|V| \cdot \diam(G))$. 


\paragraph{Proof idea}
Our proof is by direct reduction from the canonical 
$\PSPACE$-complete language of true quantified Boolean formulas.
It maps true formulas onto networks where one of the agents' posteriors
is almost entirely concentrated on $\theta = \sT$ and false formulas
onto networks where the posterior is concentrated on $\theta = \sF$.
The reduction and the proof are by induction on the number of
quantifier alternations in the Boolean formula. The base case of the induction
corresponds to such mapping for satisfiable and unsatisfiable $\3SAT$ instances.

The basic idea of the reduction is to map 
variables and clauses of the Boolean formula
onto agents or small sub-networks of agents (gadgets) in the
Bayesian network. We use other gadgets to implement some useful procedures,
like counting or logical operations.
One challenging aspect of the reduction is that, since each such operation
is implemented by Bayesian agents by broadcasting their opinions,
these ``measurements'' themselves might shift the posterior belief of
the ``observer'' agent. Therefore, we need to carefully compensate those
unintended effects at every step.

Another interesting technical aspect of the proof is related to its recursive
nature. When we establish hardness of approximation for $k$ quantifier
alternations, it means that we can place an agent  in our network such that the agent 
will be solving a ``$k$-hard'' computational problem. We then use this agent,
together with another gadget that modifies relative likelihoods of
different private signal configurations, to amplify hardness to
$k+1$ alternations.

\paragraph{Related literature}

One intriguing aspect of our result
is a connection to the Aumann's agreement theorem. There is a
well-known discrepancy
(see~\cite{CH02} for a distinctive take)
between reality, where we commonly observe (presumably) honest, well-meaning
people ``agreeing to disagree'', and the
Aumann's theorem, stating that this cannot happen for Bayesian agents
with common priors and knowledge,
i.e., the agents will always end up with the same estimate of
the state of the world after exchanging all relevant information.
Our result hints at a computational explanation, suggesting that reasonable
agreement protocols might be intractable in the presence of network structure.
This is notwithstanding some positive computational results by Hanson~\cite{Han03}
and Aaronson~\cite{Aar05}, which focus on two agents and come with their own
(perhaps unavoidable) caveats.

We find it interesting that the agents' computations in the binary action
model turn out to be not just hard, but $\PSPACE$-hard.
$\PSPACE$-hardness of partially observed Markov decision processes
(PMODPs) established by Papadimitriou and Tsitsiklis~\cite{PT87} seems to 
be a result of similar kind. On the other hand, there are clear differences:
We do not see how to implement our model as PMODP, and embedding a
$\TQBF$ instance in a PMODP looks more straightforward than
what happens in our reduction.  
Furthermore, and contrary to \cite{PT87}, we establish
hardness of approximation. 
We are not aware of many other $\PSPACE$-hardness of approximation
proofs, especially in recent years. Exceptions are results obtained
via $\PSPACE$ versions of the PCP theorem \cite{CFLS95, CFLS97}
and a few other reductions \cite{MHR94, HMS94, Jon97, Jon99} that concern, among others,
some problems on hierarchically generated graphs and an AI-motivated
problem of planning in propositional logic.

We note that there are some results on hardness of Bayesian reasoning in static
networks in the AI and cognitive science context (see \cite{Kwi18} and its
references), but this setting seems quite different from dynamic opinion
exchange models.

Finally, we observe that a natural exhaustive search algorithm for computing
the action $A(u, t)$ in the binary action model requires exponential
space (see~\cite{JMR17} for a description) and that we are not aware of
a faster, general method (but again, see~\cite{MOT16} for a polynomial time
algorithm in a variant with Gaussian signals).

\paragraph{Organization of the paper}
In Section~\ref{sec:model} we give a full description of our model,
state the results precisely and give some remarks about the proofs.
Section~\ref{sec:np} contains the proof of $\NP$-hardness,
which is then used in Section~\ref{sec:pspace} in the proof of
$\PSPACE$-hardness.
Section~\ref{sec:bounded} modifies the proof to use only a fixed
number of private signal distributions.
Section~\ref{sec:sharp} provides a proof of $\HASHP$-hardness in a related
\emph{revealed belief} model. Finally, Section~\ref{sec:conclusion} contains
some suggestions for future work.

\section{The model and our results}
\label{sec:model}

In Section~\ref{sec:model-sub} we restate the binary action model in more
precise terms and introduce some notation.
In Section~\ref{sec:results} we discuss our results in this model.
In Section~\ref{sec:belief} we define the \emph{revealed belief}
model and state a $\HASHP$-hardness result for it.
Finally, in Section~\ref{sec:proof} we explain our main proof ideas.

\subsection{Binary action model}
\label{sec:model-sub}

We consider the \emph{binary action model of Bayesian opinion exchange
  on a network}.
There is a directed graph $G = (V, E)$, the vertices of which we call
\emph{agents}.
The world has a hidden binary \emph{state} $\theta \in \{\sT, \sF\}$
with uniform prior distribution. We will analyze a process with discrete
moments $t = 0, 1, 2, \ldots$ At time $t=0$ each agent $u$
receives a private signal $S(u) \in \{0, 1\}$. The signals $S(u)$ are
random variables with distributions that are independent across agents
after conditioning on $\theta$.
Accordingly,
the distribution of $S(u)$ is
determined by its \emph{signal probabilities}
\begin{align*}
  p_{\theta_0}(u) := \Pr\left[ S(u) = 1 \mid \theta = \theta_0\right],
  \theta_0 \in \{\sT, \sF\} \; .
\end{align*}
Equivalently, it is determined by its \emph{(log)-likelihoods}
\begin{align*}
  \ell_b(u) = \ln \frac{\Pr[S(u) = b \mid \theta = \sT]}
  {\Pr[S(u) = b \mid \theta = \sF]}
  = \ln \frac{\Pr[\theta = \sT \mid S(u) = b]}{\Pr[\theta = \sF \mid S(u) = b]}
  , b \in \{0, 1\} \; .
\end{align*}
Note that there is a one-to-one correspondence between probabilities
$p_\sT$ and $p_\sF$ with $p_\sT \ne p_\sF$, and likelihoods $\ell_0$,
$\ell_1$ with
$\ell_0 \cdot \ell_1 < 0$. We will always assume
that a signal $S(u) = 1$ is evidence towards $\theta=\sT$ and vice versa.
This is equivalent to saying that $p_\sT > p_\sF$ or
$\ell_1 > 0$ and $\ell_0 < 0$.
We allow some agents to not receive private signals:
This can be ``simulated'' by giving them non-informative signals with
$p_\sT(u) = p_\sF(u)$.
We will refer to all signal probabilities taken together as the \emph{signal
  structure} of the Bayesian network.
A specific pattern of signals $s \in \{0, 1\}^{|V'|}$ (where $V'$ denotes
the subset of agents that receive informative signals)
will be called
a \emph{signal configuration}.

We assume that all this structural information is publicly known,
but the agents do not have direct access to $\theta$ or others' private signals.
Agents are presumed to be rational, to know that everyone else is rational,
to know that everyone knows, etc. (common knowledge of rationality).
At each time $t \ge 0$, we define $\mu(u, t)$ to be the
\emph{belief} of agent $u$:
The conditional probability that $\theta = \sT$ given everything that
$u$ observed at times $t' < t$.
More precisely, letting $\mathcal{N}(u)$ be the (out)neighbors of $u$
in $G$ and defining
\begin{align*}
  H(u, t) := \left\{ A(v, t'): t' < t, v \in \mathcal{N}(u) \right\} \; .
\end{align*}
as the \emph{observation history} of agent $u$ we let
\begin{align*}
  \mu(u, t) := \Pr[ \theta = \sT \mid S(u), H(u, t) ] \; .
\end{align*}
Accordingly, if $(u, v) \in E(G)$ we will say that agent $u$
\emph{observes} agent $v$.

Agent $u$ broadcasts to its in-neighbors the
\emph{action} $A(u, t) \in \{\sT, \sF\}$, which is the state of the world
that $u$ considers more likely according to $\mu(u, t)$
(assume that ties are broken in an arbitrary deterministic manner,
say, in favor of $\sF$). Then, the protocol proceeds to time step $t+1$
and the agents update their beliefs and broadcast updated actions.
The process continues indefinitely. Note that the beliefs and actions
become deterministic once the private signals are fixed.

The first two time steps of the process
are relatively easy to understand:
At time $t=0$ an agent broadcasts $A(u, 0) = \sT$ if and only if
$S(u) = 1$ and the belief $\mu(u, 1)$ can be easily computed from
the likelihood $\ell_{S(u)}(u)$. 
At time $t=1$, an agent broadcasts $A(u, 1) = \sT$ if and only if
\begin{align}
  \label{eq:09}
  \ell_{S(u)}(u) + \sum_{v \in \mathcal{N}(u)} \ell_{S(v)}(v) > 0 \; ,
\end{align}
where the private signals $S(v)$ can be inferred from
observed actions $A(v, 0)$.
The sum \eqref{eq:09} determines the likelihood associated
with belief $\mu(u, 1)$. However, at later times the actions of different neighbors
are not independent anymore and accounting for those dependencies seems
difficult.

Let $\Pi$ be a Bayesian network, i.e., a directed graph $G = (V, E)$
together with the signal structure.
We do not commit to any particular representation of probabilities
of private signals. Our reduction remains valid for any reasonable choice.
We are interested in
hardness of computing the actions that the agents need to broadcast.
More precisely, we consider complexity of computing the function
\begin{align*}
  \BAY(\Pi, t, u, S(u), H(u, t)) := A(u, t) 
\end{align*}
that computes the action $A(u, t)$ given the Bayesian network, time $t$,
agent $u$, its private signal $S(u)$ and observation history $H(u, t)$.
Relatedly, we will consider computing belief
\begin{align*}
  \BAYAB(\Pi, t, u, S(u), H(u, t)) := \mu(u, t) \; .
\end{align*}

Note that computing $\BAY$ is equivalent to distinguishing between
$\BAYAB > 1/2$ and $\BAYAB \le 1/2$.

\subsection{Our results}
\label{sec:results}

Our first result implies that computing $\BAY$ at time $t=2$ is $\NP$-hard.
We present it as a standalone theorem, since the $\NP$ reduction and
its analysis are used as a building block in the more complicated $\PSPACE$
reduction.

\begin{theorem}
  \label{thm:np}
There exists an efficient reduction from a $\3SAT$ formula $\phi$
with $N$ variables and $M$ clauses
to an input of $\BAY(\Pi, t, u, H(u, t))$ such that:
\begin{itemize}
\item The size (number of agents and edges) of Bayesian graph $G$ is $O(N + M)$,
  time is set to $t=2$ and agent $u$ does not receive a private signal.
\item All probabilities of private signals are efficiently computable real
  numbers satisfying
  \begin{align*}
    \exp(-(O(N)) \le p_{\theta_0}(v) \le 1 - \exp(-O(N)), \,
    v \in V, \, \theta_0 \in \{\sT, \sF\} \; .
  \end{align*}
\item If $\phi$ is satisfiable, then the posterior $\mu(u, 2)$ satisfies
  \begin{align*}
    \mu(u, 2) = 1 - \exp(-\Theta(N)) \; .
  \end{align*}
\item If $\phi$ is not satisfiable, then we have
  \begin{align*}
    \mu(u, 2) = \exp(-\Theta(N)) \; .
  \end{align*}
\end{itemize}
\end{theorem}

\begin{corollary}
Both
distinguishing between $\BAYAB >1-\exp(-O(N))$
and $\BAYAB < \exp(-O(N))$ and computing $\BAY$ are $\NP$-hard.
\end{corollary}

Our main result improves Theorem~\ref{thm:np} to $\PSPACE$-hardness.
It is a direct reduction from the canonical $\PSPACE$-complete language
of true quantified Boolean formulas $\TQBF$.

\begin{theorem}
  \label{thm:pspace}
There exists an efficient reduction from a $\TQBF$ formula $\Phi$
\begin{align*}
  \Phi = Q_Kx_K \cdots \exists x_1 \phi(x_K, \ldots, x_1) \; ,
\end{align*}
where $\phi$ is a 3-CNF formula with $N$ variables
and $M$ clauses, there are $K$ variable blocks with alternating quantifiers
and the last quantifier is existential,
to an input of $\BAY(\Pi, t, u, H(u, t))$ such that:
\begin{itemize}
\item The number of agents in Bayesian graph $G$ is $O(N^2(N+M))$,
  time is set to $t = 2K$ and agent $u$ does not receive a private signal.
\item All probabilities of private signals are efficiently computable real
  numbers satisfying
  \begin{align}
    \label{eq:21}
    \exp(-(O(N)) \le p_{\theta_0}(v) \le 1 - \exp(-O(N)), \,
    v \in V, \, \theta_0 \in \{\sT, \sF\} \; .
  \end{align}
\item If $\Phi$ is true, then $\mu(u, 2K) = 1 - \exp(-\Theta(N))$.
  If $\Phi$ is false, then $\mu(u, 2K) = \exp(-\Theta(N))$.
\end{itemize}
\end{theorem}

\begin{corollary}
Both
distinguishing between $\BAYAB > 1-\exp(-O(N))$
and $\BAYAB < \exp(-O(N))$ and computing $\BAY$ are $\PSPACE$-hard.
\end{corollary}

\begin{remark}
  Note that the statement of Theorem~\ref{thm:pspace}
  immediately gives $\Sigma_K$- and $\Pi_K$-hardness of
  approximating $\BAYAB$ at time $t=2K$.
\end{remark}

\begin{remark}
  For ease of exposition we define networks in the reductions
  to be directed, but due to additional structure that we impose
  (see paragraph ``Network structure and significant times''
  in Section~\ref{sec:np}) it is easy to see that they can be assumed
  to be undirected.
  This is relevant insofar as a strong form of learning occurs only on undirected
  graphs (see~\cite{MST14} for details).
\end{remark}

One possible objection to Theorem~\ref{thm:pspace} is that it uses signal
distributions with probabilities exponentially close to zero and one.
We do not think this is a significant issue, and it
helps avoid some technicalities. Nevertheless,
in Section~\ref{sec:bounded} we prove a version of Theorem~\ref{thm:pspace}
where all private signals come from a fixed family of say,
at most fifty distributions. This is at the cost of (non-asymptotic) increase
in the size of the graph.

\begin{theorem}
  \label{thm:bounded}
  The reduction from Theorem~\ref{thm:pspace} can be modified such that
  all private signals come from a fixed family of at most
  fifty distributions.
\end{theorem}

\begin{remark}
  It is possible to modify our proofs to give hardness
  of distinguishing between $\mu(u, t) \le \exp(-O(N^K))$
  and $\mu(u, t) \ge 1-\exp(-O(N^K))$ for any constant $K$
  (recall that $N$ is the number of variables in the formula $\phi$).
  This is at the cost of allowing signal probabilities in the range
  \begin{align*}
    \exp(-(O(N^K)) \le p_{\theta_0}(v) \le 1 - \exp(-O(N^K))
  \end{align*}
  or, in the bounded signal case, increasing the network size to
  $O(N^{K+2})$. Consequently, in the latter case we get hardness of approximation
  up to $\exp(-O(|V|^{\alpha}))$ factor for any constant $\alpha < 1$,
  where $|V|$ is the number of agents.
\end{remark}

\subsection{Revealed beliefs}
\label{sec:belief}

In a natural variant of our model the agents
act in exactly the same manner, except that they reveal their
full beliefs
$A(u, t) = \mu(u, t)$ rather than just estimates of the state $\theta$.
Accordingly, we call it the 
\emph{revealed belief} model.
We suspect that binary action
and revealed belief models have similar computational powers.
Furthermore, we conjecture that if the agents broadcast their beliefs
rounded to a (fixed in advance) polynomial number of significant digits,
then our techniques can be extended to establish a similar
$\PSPACE$-hardness result.

However, if one instead assumes that the beliefs are broadcast up to an arbitrary
precision,
our proof fails for a rather
annoying reason: When implementing alternation from $\NP$ to $\Pi_2$
in the binary action model, if a formula $\phi$ has no satisfying assignments,
we can exactly compute the belief of the $\NP$ observer agent.
However, in case $\phi$ has a satisfying assignment, we can compute the belief
only with high, but imperfect precision.
The reason is that the exact value of the belief depends on the number of
satisfying assignments of $\phi$.
This imperfection can be ``rounded away'' if the agents output a discrete guess
for $\theta$, but we do not know how to handle it if the beliefs are
broadcast exactly.

Nevertheless, in Section~\ref{sec:sharp} we present a
$\HASHP$-hardness proof in the revealed belief model.
The proof is by reduction from counting satisfying assignments in a
$\2SAT$ formula.
However, since the differences in the posterior corresponding to
different numbers of satisfying assignments are small,
it is not clear if they can be amplified, and consequently we do not demonstrate
hardness of approximation (similar as in~\cite{PT87}).
For ease of exposition we introduce an additional relaxation to the model by
allowing some agents to receive ternary private signals.

\begin{theorem}
  \label{thm:sharp}
  Assume the revealed belief model with beliefs transmitted up to
  arbitrary precision
  and call the respective computational problem $\BAYAB$.
  Additionally, assume that some agents receive ternary
  signals $S(u) \in \{0, 1, 2\}$.
  
  There exists an efficient reduction that maps
  a $\2SAT$ formula $\phi$ with $N$ variables, $M$ clauses
  and $A$ satisfying assignments
  to an instance of $\BAYAB(\Pi, t, u, H(u, t))$ such that:
  \begin{itemize}
  \item Bayesian network $G$ has size $O(N+M)$, time is set to $t=2$
    and agent $u$ does not receive a private signal.
  \item All private signal probabilities come from a fixed family of at
    most ten distributions.
  \item
  The likelihood of $u$ at time $t=2$
  satisfies
  \begin{align*}
    \frac{A}{2^N} \left(1 - \frac{1}{4^N}\right)
    \le \frac{\mu(u, 2)}{1-\mu(u, 2)} \le
    \frac{A}{2^N}\left(1+\frac{1}{4^N}\right) \; .
  \end{align*}
  In particular, rounding this likelihood to the nearest multiple
  of ${2^{-N}}$ yields $A\cdot 2^{-N}$ and allows to recover $A$.
  \end{itemize}
\end{theorem}

\subsection{Main proof ideas}
\label{sec:proof}
The NP-hardness proof (in Section~\ref{sec:np})
uses an analysis of a composition of several gadgets.
We will think of the agent $u$ from input to $\BAY$ as
``observer'' and accordingly call it $\OBS$.
The Bayesian network features
gadgets that represent variables and clauses.
The private signals in variable gadgets
correspond to assignments $x$ to the formula $\phi$.
Furthermore, there is an ``evaluation agent'' $\EVAL$ that interacts
with all clause gadgets.
We use more gadgets that ``implement'' counting to ensure that
what $\OBS$ observes is consistent with one of two possible kinds
of signal configurations:
\begin{itemize}
\item  $S(\EVAL) = 1$ and the signals of variable agents correspond to an
  arbitrary assignment $x$.
\item $S(\EVAL) = 0$ and the signals of variable agents correspond to a
  \emph{satisfying} assignment $x$.
\end{itemize}
Then, we use another gadget to ``amplify'' the information that is conveyed
about the state of the world by the signal $S(\EVAL)$. If $\phi$ has
no satisfying assignment, then $S(\EVAL) = 1$ and this becomes amplified
to a near-certainty that $\theta = \sF$ (for technical reasons this is
the opposite conclusion than suggested by $S(\EVAL) = 1$).
On the other hand, we design the signal structure such that even a single
satisfying assignment tips the scales and amplifies to $\theta = \sT$ with high probability (whp).

We note that one technical challenge in executing this plan is that
some of our gadgets are designed to ``measure'' (e.g., count) certain
properties of the network, but these measurements use auxiliary agents with
their own private signals, affecting Bayesian posteriors. We need to be careful to
cancel out these unintended effects at every step.

The high-level idea to improve on the $\NP$-hardness proof is that
once we know that agents can solve hard problems, we can use them
to help the observer agent solve an even harder problem.
Of course this has to be done in a careful way, since the answer to
a partial problem cannot be directly revealed to the observer
(the whole point is that we do not know a priori what this answer is).

The $\PSPACE$ reduction is defined and Theorem~\ref{thm:pspace} proved
by induction.
The base case is the Bayesian network
from Theorem~\ref{thm:np}, but with observer agent
directly observing private signals in the first $K-1$ variable blocks.
Then, we proceed
to add ``intermediate observers'', each of them observing one
variable block less, and
interacting via a gadget with the previous observer to implement the quantifier
alternation by adjusting likelihoods of different assignments to
variables in $\Phi$.

It is useful to view $\Phi$
as a game where two players set quantified variables (proceeding left-to-right)
in $\Phi$. One player sets variables under existential quantifiers with
the objective of evaluating the $3$-CNF formula $\phi$ true.
The other player sets variables under universal quantifiers with the
objective of evaluating $\phi$ false.
Under that interpretation,
our reduction has the following property: The final observer
agent $\OBS$ concludes that the assignment $x$ formed by private signals
with high probability corresponds to a ``game transcript'' in the game
played according to a winning strategy. Depending on which player has
the winning strategy, the state of the world is either $\sT$ or $\sF$
whp.

\section{\texorpdfstring{$\NP$}{NP}-hardness:
  Proof of  Theorem~\ref{thm:np}}
\label{sec:np}

We start with the $\NP$-hardness result by reduction from $\3SAT$. The reduction
is used as a building block in the $\PSPACE$-hardness proof, but it is also
useful in terms of developing intuition for the more technical proof of
Theorem~\ref{thm:pspace}. We proceed by explaining gadgets that we
use, describing how to put them together in the full reduction and
proving the correctness.

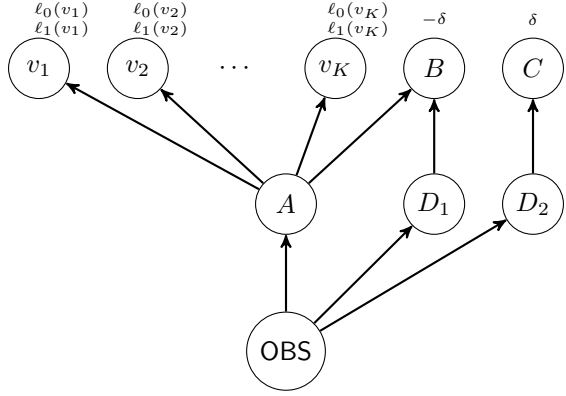
\begin{figure}[ht]\centering
  \caption{Threshold gadget.}
  \label{fig:threshold-gadget}
  \begin{tikzpicture}
    \node [draw, circle, label=center:$v_1$] (v1) at (0, 0) {\phantom{$v_K$}};
    \node [font=\fontsize{6}{0}\selectfont, align=center] at (0.3, 0.65) {$\ell_0(v_1)$\\$\ell_1(v_1)$};
    \node [draw, circle, label=center:$v_2$] (v2) at (1.3, 0) {\phantom{$v_K$}};
    \node [font=\fontsize{6}{0}\selectfont, align=center] at (1.6, 0.65)
    {$\ell_0(v_2)$\\$\ell_1(v_2)$};
    \node (dots) at (2.6, 0) {$\cdots$};
    \node [draw, circle, label=center:$v_K$] (vK) at (3.9, 0) {\phantom{$v_K$}};
    \node [font=\fontsize{6}{0}\selectfont, align=center] at (4.2, 0.65)
    {$\ell_0(v_K)$\\$\ell_1(v_K)$};
    \node [draw, circle, label=center:$B$] (b) at (5.2, 0) {\phantom{$v_K$}};
    \node [font=\fontsize{6}{0}\selectfont] at (5.2, .65) {$-\delta$};
    \node [draw, circle, label=center:$C$] (c) at (6.5, 0) {\phantom{$v_K$}};
    \node [font=\fontsize{6}{0}\selectfont] at (6.5, .65) {$\delta$};
    
    \node [draw, circle, label=center:$A$] (a) at (3.25, -1.8) {\phantom{$v_K$}};
    \node [draw, circle, label=center:$D_1$] (dummy1) at (5.2, -1.8) {\phantom{$v_K$}};
    \node [draw, circle, label=center:$D_2$] (dummy2) at (6.5, -1.8) {\phantom{$v_K$}};
    \node [draw, circle] (OBS) at (3.25, -3.75) {$\OBS$};

    \draw [myArrow] (a) -- (v1);
    \draw [myArrow] (a) -- (v2);
    \draw [myArrow] (a) -- (vK);
    \draw [myArrow] (a) -- (b);
    \draw [myArrow] (dummy1) -- (b);
    \draw [myArrow] (dummy2) -- (c);
    \draw [myArrow] (OBS) -- (a);
    \draw [myArrow] (OBS) -- (dummy1);
    \draw [myArrow] (OBS) -- (dummy2);
  \end{tikzpicture}
\end{figure}

\paragraph{Threshold gadget}
Say there are agents $v_1, \ldots, v_K$ that do not observe
anyone and receive private signals
$S(v_i)$ with respective likelihoods $\ell_0(v_i)$ and $\ell_1(v_i)$.
Additionally, there is an observer agent $\OBS$ and we would like
to reveal to it, at time $t = 1$, that the sum of likelihoods of agents
$v_1, \ldots, v_K$ exceeds some threshold $\delta$:
\begin{align*}
  L := \sum_{i=1}^K \ell_{S(v_i)}(v_i) > \delta \; ,
\end{align*}
without disclosing
anything else about the private signals.\footnote{
  We assume that $\delta$ is chosen such that 
  $L = \delta$ never happens.
}
This is achieved by the gadget in Figure~\ref{fig:threshold-gadget}.

We describe the gadget for $\delta > 0$.
Agent $B$ receives private signal with
$\ell_0(B) = -\delta$ (and arbitrary $\ell_1(B)$)
and agent $C$ with $\ell_1(C) = \delta$.
Agents $A$, $D_1$ and $D_2$ (we will call the latter two the ``dummy'' agents)
do not receive private signals.
Our overall reduction will demonstrate the hardness of computation for
agent $\OBS$. Therefore, we need to specify the observation history of $\OBS$.
By our tie-breaking convention, it must be
$A(A, 0) = A(D_1, 0) = A(D_2, 0) = \sF$. Furthermore, we specify
$A(A, 1) = A(D_2, 1) = \sT$ and $A(D_1, 1) = \sF$.

Based on that information,
agent $\OBS$ can infer that $S(B) = 0$, $S(C) = 1$
and, since the action $A(A, 2)$ is determined by the sign of $L-\delta$,
that $L > \delta$.
The purpose of agent $C$ is to counteract the effect of this ``measurement''
on the estimate of the state of the world by $\OBS$. More precisely,
let
\begin{align}
  \label{eq:10}
  P(s_1, \ldots, s_K, \theta_0)
  &:=
    \Pr\left[ \bigwedge_{i=1}^K S(v_i) = s_i \land \theta = \theta_0
    \right] \; ,\\
  P(s_1, \ldots, s_K, s_B, s_C, \theta_0)
  & :=
    \Pr\left[ \bigwedge_{i=1}^KS(v_i)=s_i\land S(B)=s_B\land
    S(C)=s_C \land \theta=\theta_0\right] \; .
\end{align}
Based on the discussion above, we have the following:

\begin{claim}
  \label{cl:threshold}
  Let $s_1, \ldots, s_K$ be private signals of $v_1, \ldots, v_K$.
  Similarly, let $(s_B, s_C)$ be private signals of $B$ and $C$.
  Then:
  \begin{itemize}
  \item
    If $\sum_{i=1}^K \ell_{s_i}(v_i) < \alpha$, then there are no $(s_B, s_C)$
    that make $(s_1, \ldots, s_K, s_B, s_C)$
    consistent with observations of $\OBS$.
  \item 
    If $\sum_{i=1}^K \ell_{s_i}(v_i) > \alpha$, then there exists unique
    configuration $(s_B, s_C)$  consistent
    with observations of $\OBS$ and the (prior) probability of this
    configuration when the state is $\theta_0$ is
    \begin{align}
      \label{eq:11}
      P(s_1, \ldots, s_K, s_B, s_C, \theta_0)
      = P(s_1, \ldots, s_K, \theta_0) \cdot \alpha \; ,
    \end{align}
    where $\alpha := (1-p_\sT(B))p_\sT(C) = e^{\ell_0(B)  + \ell_1(C) }(1-p_\sF(B))p_\sF(C) = (1-p_\sF(B))p_\sF(C)$
    does not depend on $\theta_0$.
  \end{itemize}
\end{claim}
  
\begin{figure}[ht]\centering
  \caption{Notation for the threshold gadget.}
  \label{fig:threshold-symbol}
  \begin{tikzpicture}
    \node [draw, circle, label=center:$v_1$] (v1) at (0, 0) {\phantom{$v_K$}};
    \node [draw, circle, label=center:$v_2$] (v2) at (1.3, 0) {\phantom{$v_K$}};
    \node (dots) at (2.6, 0) {$\cdots$};
    \node [draw, circle, label=center:$v_K$] (vK) at (3.9, 0) {\phantom{$v_K$}};
    \coordinate (top) at (1.5, -1);
    \draw (0.8, -1) -- (2.2, -1)  -- (2.2, -2) -- (0.8, -2) -- cycle;
    \draw [myArrow] (top) -- (v1);
    \draw [myArrow] (top) -- (v2);
    \draw [myArrow] (top) -- (vK);
    \node [font=\fontsize{15}{0}\selectfont] at (1.5, -1.5) {$>\delta$};
    \node [draw, circle] (OBS) at (1.5, -3.4) {$\OBS$};
    \draw [myArrow] (OBS) -- (1.5, -2);
  \end{tikzpicture}
\end{figure}
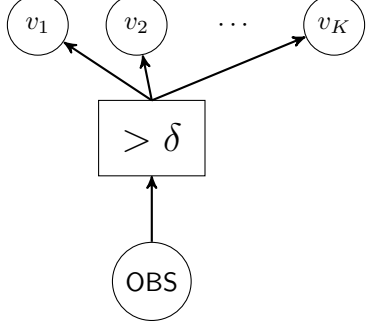

Similar reasoning can be made for the case when $\delta < 0$
and/or checking the opposite inequality $L < \delta$.
We will say that an agent $\OBS$ observes a threshold gadget if it
observes agents $A$, $D_1$ and $D_2$ and denote it as shown in
Figure~\ref{fig:threshold-symbol}. Note that in our diagrams we use circles to
denote agents and boxes to denote gadgets.
The latter typically contain several auxiliary agents.

\paragraph{Network structure and significant  times}
It might appear that the threshold gadget
is more complicated than needed. The reason for this is that we will impose
certain additional structure on the graph to facilitate its analysis, and later use it in the
proof of Theorem~\ref{thm:pspace}.
Specifically, we will always make sure that the graph is a DAG,
with only the observer agent having in-degree zero. All agents
that receive private signals will have out-degree zero, and all agents with
non-zero out-degree will not receive private signals (recall a directed edge $A \to B$ indicates that $A$ observes $B$).

Furthermore, we will arrange the graph such that each agent 
will learn new information at a single, fixed time step.
That is, for every agent $A$ there will exist a \emph{significant time} $t(A)$
such that $\mu(A, t) = 1/2$ for $t' < t(A)$ and $\mu(A, t') = \mu(A, t(A))$
for $t' > t(A)$. If $A$ receives a private signal, then $t(A) = 0$.
Otherwise, $t(A)$ is determined by the (unique) path length
from $A$ to an out-degree zero agent.
For example, in Figure~\ref{fig:threshold-gadget} significant times
are $t(A) = t(D_1) = t(D_2) = 1$ and $t(\OBS) = 2$.

Accordingly, we will use notation $\mu(A)$ and $A(A)$ to denote agent
beliefs and actions at the significant time. Let $A$ and $B$ be agents
with $t(A) < t(B)-1$. In the following, we will sometimes say that $B$
observes $A$, even though that would contradict the significant time
requirement (a direct edge $B \to A$ implies that  $t(A) = t(B)-1$). Whenever we do so, it should be understood that there is
a path of ``dummy'' nodes of appropriate length between $A$ and $B$
(cf.~$D_1$ and $D_2$ in Figure~\ref{fig:threshold-gadget}). For clarity, we will
omit dummy nodes from the figures. 

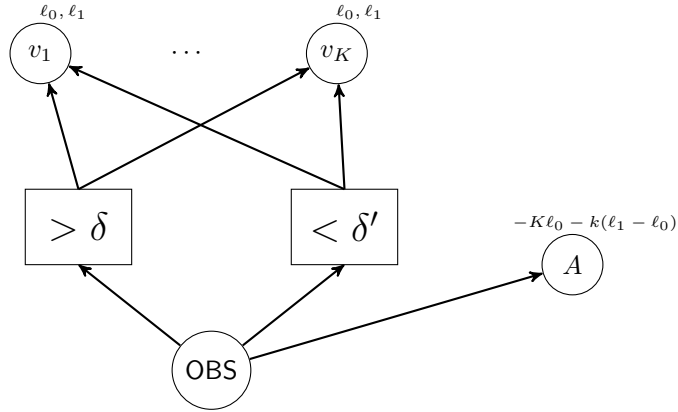
\begin{figure}[ht]\centering
  \caption{Counting gadget.}
  \label{fig:counting-gadget}
  \begin{tikzpicture}
    \node [draw, circle, label=center:$v_1$] (v1) at (0, 0) {\phantom{$v_K$}};
    \node [font=\fontsize{6}{0}\selectfont, align=center] at (0.3, 0.55)
    {$\ell_0, \ell_1$};    
    \node (dots) at (1.95, 0) {$\cdots$};
    \node [draw, circle, label=center:$v_K$] (vK) at (3.9, 0) {\phantom{$v_K$}};
    \node [font=\fontsize{6}{0}\selectfont, align=center] at (4.2, 0.55)
    {$\ell_0, \ell_1$};    
    \coordinate (top) at (0.5, -1.8);
    \draw  (-0.2, -1.8) -- (1.2, -1.8)  -- (1.2, -2.8) -- (-0.2, -2.8) -- cycle;
    \draw [myArrow] (top) -- (v1);
    \draw [myArrow] (top) -- (vK);
    \node [font=\fontsize{15}{0}\selectfont] at (0.5, -2.3) {$>\delta$};
    \coordinate (top2) at (4, -1.8);
    \draw  (3.3, -1.8) -- (4.7, -1.8) -- (4.7, -2.8) -- (3.3, -2.8) -- cycle;
    \draw [myArrow] (top2) -- (v1);
    \draw [myArrow] (top2) -- (vK);
    \node [font=\fontsize{15}{0}\selectfont] at (4, -2.3) {$<\delta'$};
    \node [draw, circle, label=center:$A$] (A) at (7, -2.8) {\phantom{$v_K$}};
    \node [font=\fontsize{6}{0}\selectfont, align=center] at (7.3, -2.25)
    {$-K\ell_0-k(\ell_1-\ell_0)$};    
    \node [draw, circle] (OBS) at (2.25, -4.2) {$\OBS$};
    \draw [myArrow] (OBS) -- (0.5, -2.8);
    \draw [myArrow] (OBS) -- (4, -2.8);
    \draw [myArrow] (OBS) -- (A);
  \end{tikzpicture}
\end{figure}

\paragraph{Counting gadget}
Assume now that the agents $v_1, \ldots, v_K$ receive private signals
with identical likelihoods $\ell_0 < 0$ and $\ell_1 > 0$ and that a number
$k$, $1 \le k \le K$ is given.
Then, building on the threshold gadget, it is easy to convey the information
that exactly $k$ out of $K$ agents received private signal $1$.
Letting $\delta := K\ell_0 + (k-0.5)(\ell_1-\ell_0)$ and
$\delta' := \delta + \ell_1 - \ell_0$, we compose two threshold gadgets
as shown in Figure~\ref{fig:counting-gadget}.

Agent $A$ is optional: Depending on our needs we will use the counting gadget
with or without it. It is used to preserve the original belief of $\OBS$
after learning the count of private signals of agents $v_i$. It receives
a private signal with $\ell_b(A) := \ell := -K\ell_0 - k(\ell_1-\ell_0)$
for appropriate $b$ (depending on the sign of $\ell$) and broadcasts
the corresponding state $\theta_0$. By similar analysis
as for the threshold gadget and using the $P(\cdot)$ notation as in
$\eqref{eq:10}$--$\eqref{eq:11}$ we have:

\begin{claim}
  \label{cl:counting}
  Let $s_1, \ldots, s_K$ be private signals of agents $v_1, \ldots, v_K$.
  Let $s$ represent private signals of all auxiliary agents in the threshold
  gadgets and $s_A$ a private signal of agent $A$.
  
  Then, the only configurations $s_1, \ldots, s_K$  consistent with
  observations of $\OBS$ are those for which
  $\sum_{i=1}^K s_i = k$. Furthermore, for any such configuration
  there exists a unique configuration $s$ (and $s_A$, if agent $A$ is present)
  such that (depending on the presence of $A$):
\begin{align*}
  P(s_1, \ldots, s_K, s, \theta_0)
  &= P(s_1, \ldots, s_K, \theta_0) \cdot \alpha =
    P(\theta_0) \cdot \alpha \;,\\
  P(s_1, \ldots, s_K, s, s_A, \theta_0) &= \beta \; ,
\end{align*}
where $\alpha := \alpha(k, K, \ell_0, \ell_1) > 0$ is easily computable and
does not depend
on $s_1, \ldots, s_K$ or $\theta_0$, but the value of the other term
$P(\theta_0)$ is in general dependent on $\theta_0$.
On the other hand, if $A$ is present, then
$\beta := \beta (k, K, \ell_0, \ell_1) > 0$ does not depend at all on
private signals or state of the world.
\end{claim}

\begin{figure}[ht]\centering
  \caption{Two counting gadgets illustrating the notation.
    The left-hand gadget ensures that at least $k$ agents received ones.
    The right-hand gadget ensures that exactly $k$ agents received ones.
    Furthermore, the equivalence symbol on the
    right-hand gadget denotes presence of the optional agent $A$.}
  \label{fig:counting-symbol}
  \begin{tikzpicture}
    \node [draw, circle, label=center:$u_1$] (u1) at (0, 0) {\phantom{$v_K$}};
    \node (dots) at (1.1, 0) {$\cdots$};
    \node [draw, circle, label=center:$u_K$] (uK) at (2.2, 0) {\phantom{$v_K$}};
    \draw (0.4, -0.8)  -- (1.8, -0.8) -- (1.8, -2.1) -- (0.4, -2.1) -- cycle;
    \node [font=\fontsize{15}{0}\selectfont] at (1.1, -1.3) {$\ge k$};
    
    \node [draw, circle, label=center:$v_1$] (v1) at (4, 0) {\phantom{$v_K$}};
    \node (dots) at (5.1, 0) {$\cdots$};
    \node [draw, circle, label=center:$v_K$] (vK) at (6.2, 0) {\phantom{$v_K$}};
    \draw (4.4, -0.8)  -- (5.8, -0.8) -- (5.8, -2.1) -- (4.4, -2.1) -- cycle;
    \node [font=\fontsize{15}{0}\selectfont] at (5.1, -1.3) {$=k$};
    \node [font=\fontsize{15}{0}\selectfont] at (5.5, -1.8) {$\equiv$};
    
    \node [draw, circle] (OBS) at (3.1, -3.5) {$\OBS$};
    \draw [myArrow] (OBS) -- (1.1, -2.1);
    \draw [myArrow] (OBS) -- (5.1, -2.1);
    \draw [myArrow] (0.5, -0.8) -- (u1);
    \draw [myArrow] (1.7, -0.8) -- (uK);
    \draw [myArrow] (4.5, -0.8) -- (v1);
    \draw [myArrow] (5.7, -0.8) -- (vK);
  \end{tikzpicture}
\end{figure}
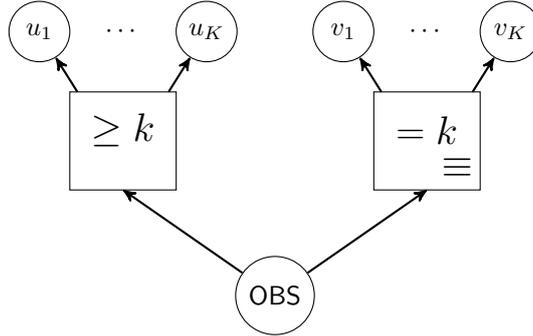

If agent $A$ is omitted, the same technique can be used to obtain
inequalities (e.g., checking that at least $k$ out of $K$
private signals are ones).
We will say that an agent $\OBS$ observes the counting gadget if it observes
both respective threshold gadgets (and $A$, if present). We will denote 
counting gadgets as in Figure~\ref{fig:counting-symbol}.

\begin{figure}[ht]\centering
  \caption{Notation for not-equal gadget.}
  \label{fig:not-equal}
  \begin{tikzpicture}
    \node [draw, circle, label=center:$u$] (u) at (0, 0) {\phantom{$u_K$}};
    \node [draw, circle, label=center:$v$] (v) at (1.5, 0) {\phantom{$u_K$}};
    \node [font=\fontsize{15}{0}\selectfont] at (0.75, -1.3) {$\ne$};
    \draw (0.25, -0.8) -- (1.25, -0.8) -- (1.25, -1.8) -- (0.25, -1.8) -- cycle;
    \draw [myArrow] (0.4, -0.8) -- (u);
    \draw [myArrow] (1.1, -0.8) -- (v);
    \node [draw, circle] (OBS) at (0.75, -3) {$\OBS$};
    \draw [myArrow] (OBS) -- (0.75, -1.8);
  \end{tikzpicture}
\end{figure}
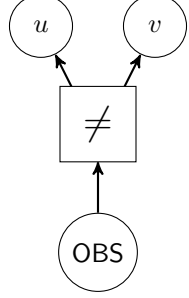

\paragraph{Not-equal gadget} Another related gadget that we will use 
reveals to the observer that two agents $u, v$ with likelihoods
$\ell_0, \ell_1$ and $m_0, m_1$, respectively,
receive opposite signals $S(u) \ne S(v)$.
Since $\ell_0 < \ell_1$ and $m_0 < m_1$,
this is achieved by using two threshold gadgets to check that
\begin{align*}
\ell_0 + m_0 <  \ell_{S(u)} + m_{S(v)} < \ell_1 + m_1 \; ,
\end{align*}
where we set the thresholds in the threshold gadgets as
$\ell_0 + m_0 + \eps$ and $\ell_1 + m_1 - \eps$ for an appropriately small
$\eps > 0$. We will denote the not-equal gadget as in Figure~\ref{fig:not-equal}.

\paragraph{Variable and clause gadgets}
Our reduction is from the standard form of $\3SAT$,
where we are given a CNF formula on $N$ variables $x_1, \ldots, x_N$.
The formula is a conjunction of $M$ clauses $C_1, \ldots, C_M$, where each clause
is a disjunction of exactly three literals on distinct variables.

We introduce two global agents. One of them is called $\OBS$ and we mean it
as an ``observer agent''. This is the agent for which we establish hardness
of computation. We will follow the rule that $\OBS$ observes all gadgets
that are present in the network. Second, we place an ``evaluation agent''
$\EVAL$ with private signals $p_\sT := 0.9$ and $p_\sF := 0.4$.

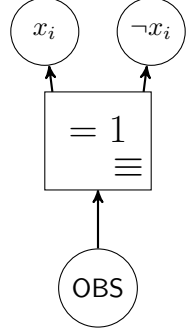
\begin{figure}[ht]\centering
  \caption{Variable gadget.}
  \label{fig:variable-gadget}
  \begin{tikzpicture}
    \node [draw, circle, label=center:$x_i$] (xi) at (0, 0) {\phantom{$\lnot x_i$}};
    \node [draw, circle] (notxi) at (1.4, 0) {$\lnot x_i$};
    \draw (0, -0.8)  -- (1.4, -0.8) -- (1.4, -2.1) -- (0, -2.1) -- cycle;
    \node [font=\fontsize{15}{0}\selectfont] at (0.7, -1.3) {$=1$};
    \node [font=\fontsize{15}{0}\selectfont] at (1.1, -1.8) {$\equiv$};
    \node [draw, circle] (OBS) at (0.7, -3.4) {$\OBS$};

    \draw [myArrow] (OBS) -- (0.7, -2.1);
    \draw [myArrow] (0.1, -0.8) -- (xi);
    \draw [myArrow] (1.3, -0.8) -- (notxi);
  \end{tikzpicture}
\end{figure}

Furthermore, for each variable in the CNF formula,
we introduce two agents $x_i$ and $\lnot x_i$ that
receive private signals given by $p_\sT$ and $p_\sF$. Then, we
encompass those two agents in a counting gadget as shown in
Figure~\ref{fig:variable-gadget}.

Then, for each clause $C_i$, we introduce
a counting gadget on four agents:
Three agents corresponding to the literals in the clause (note that they
are observed directly and not through the variable gadgets), and
the $\EVAL$ agent. The gadget ensures that at least one of those agents
received signal $1$.
Illustration is provided in Figure~\ref{fig:clause-gadget}.

\begin{figure}[ht]\centering
  \caption{Clause gadget.}
  \label{fig:clause-gadget}
  \begin{tikzpicture}
    \node [font=\fontsize{15}{0}\selectfont] at (0, 0)
    {$C_i = \lnot x_1 \lor x_3 \lor \lnot x_N$};

    \node [draw, circle, label=center:$\lnot x_1$] (x1) at (-2.25, -1.2)
    {\phantom{$\lnot x_n$}};
    \node [draw, circle, label=center:$x_3$] (x3) at (-0.75, -1.2)
    {\phantom{$\lnot x_n$}};
    \node [draw, circle, label=center:$\lnot x_N$] (xn) at (0.75, -1.2)
    {\phantom{$\lnot x_n$}};
    \node [draw, ellipse] (EVAL) at (2.25, -1.2) {$\EVAL$};
    \node [draw, circle] (OBS) at (0, -6) {$\OBS$};

    \draw (-1, -3) -- (1, -3) -- (1, -4.5) -- (-1, -4.5) -- cycle;
    \node [font=\fontsize{15}{0}\selectfont] at (0, -3.75) {$\ge1$};

    \draw [myArrow] (-0.5, -3) to [bend left] (x1);
    \draw [myArrow] (-0.2, -3) to (x3);
    \draw [myArrow] (0.1, -3) to (xn);
    \draw [myArrow] (0.5, -3) to [bend right] (EVAL);
    \draw [myArrow] (OBS) to (0, -4.5);
  \end{tikzpicture}
\end{figure}  

\paragraph{The reduction} We put the agents $\EVAL$ and $\OBS$ and the variable
and clause gadgets together, as explained in previous paragraphs.
Finally, we add two more agents $A$ and $B$. We will choose
a natural number $b := C\cdot N$ for an absolute big enough constant $C > 0$.
Agent $A$ receives private signals with $p_\sT(A) = 1-\alpha_1^b$
and $p_\sF(A) = \alpha_2^b$ and agent $B$ with $p_\sT(B) = 1-\alpha_3^b$
and $p_\sF(B) = \alpha_4^b$ for some $\alpha_1, \ldots, \alpha_4$ that
we will choose shortly. Let the corresponding likelihoods
be $o_1, o_2, o_3, o_4$ (note that $o_1, o_3 > 0$ and $o_2, o_4 < 0$).
We also insert two not-equal gadgets observed by $\OBS$:
One of them is put between $\EVAL$ and $A$ and the other one
between $A$ and $B$.
The overall construction is illustrated in
Figure~\ref{fig:sat-reduction}.

We are reducing to the problem of computing the action of agent
$\OBS$ at its significant time $t=2$.
Note that $\OBS$ observes all gadgets in the graph,
and only gadgets. In particular, $\OBS$ directly infers the signals of all auxiliary agents in the gadgets,
but the same cannot be said about the private signals of variable agents.
The observation history $H(OBS, 2)$ is naturally determined by specifications
of the gadgets.

\begin{figure}[ht]\centering
  \caption{$\3SAT$ reduction for $\phi(x) = C_1 \land \ldots \land C_M$.}
  \label{fig:sat-reduction}
  \begin{tikzpicture}
    \draw (0, 10) -- (1, 10) -- (1, 9) -- (0, 9) -- cycle;
    \node at (0.5, 9.5) {$x_1$};

    \node at (3, 9.5) {$\cdots$};

    \draw (5, 10) -- (6, 10) -- (6, 9) -- (5, 9) -- cycle;
    \node at (5.5, 9.5) {$x_N$};

    \draw (0.5, 7.5) -- (1.5, 7.5) -- (1.5, 6.5) -- (0.5, 6.5) -- cycle;
    \node at (1, 7) {$C_1$};

    \node at (3, 7) {$\cdots$};
    
    \draw (4.5, 7.5) -- (5.5, 7.5) -- (5.5, 6.5) -- (4.5, 6.5) -- cycle;
    \node at (5, 7) {$C_M$};

    \node [draw, ellipse] (EVAL) at (7.5, 9.25) {$\EVAL$};
    \node [font=\fontsize{6}{0}\selectfont] at (7.8, 9.75) {$\ell_1, \ell_0$};
    \node [draw, circle] (A) at (9.5, 9.25) {$A$};
    \node [font=\fontsize{6}{0}\selectfont] at (9.8, 9.75) {$o_1, o_2$};
    \node [draw, circle] (B) at (11, 9.25) {$B$};
    \node [font=\fontsize{6}{0}\selectfont] at (11.3, 9.75) {$o_3, o_4$};

    \node [font=\fontsize{15}{0}\selectfont] at (8.5, 8) {$\ne$};
    \draw (7.65, 8.5) -- (9.35, 8.5) -- (9.35, 7.5) -- (7.65, 7.5) -- cycle;
    
    \node [font=\fontsize{15}{0}\selectfont] at (10.5, 8) {$\ne$};
    \draw (9.65, 8.5) -- (11.35, 8.5) -- (11.35, 7.5) -- (9.65, 7.5) -- cycle;

    \node [draw, circle] (OBS) at (5.5, 5) {$\OBS$};

    \draw [myArrow] (OBS) -- (1, 6.5);
    \draw [myArrow] (OBS) -- (5, 6.5);
    \draw [myArrow] (OBS.west) to [out=210, in=315] (-1, 6)
    to [out=135, in=225] (0.5, 9);
    \draw [myArrow] (OBS.north east) to [bend right] (5.5, 9);
    \draw [myArrow] (OBS) to (8.5, 7.5);
    \draw [myArrow] (OBS) to (10.5, 7.5);
    
    \draw [myArrow] (1.5, 7.5) to [bend right=20] (EVAL);
    \draw [myArrow] (5.5, 7.5) to (EVAL.south);
    \draw [myArrow] (4.7, 7.5) to (4, 9.8);
    \draw [myArrow] (1, 7.5) to (1.8, 9.8);

    \draw [myArrow] (8.5, 8.5) to (EVAL.south east);
    \draw [myArrow] (8.7, 8.5) to (A);
    \draw [myArrow] (10.3, 8.5) to (A);
    \draw [myArrow] (10.5, 8.5) to (B);
  \end{tikzpicture}
\end{figure}  

\paragraph{Analysis}
As a preliminary matter, the reduction indeed produces an
instance of polynomial size: The size of the graph is $O(N + M)$
and the probabilities of private signals satisfy
\begin{align*}
  \exp(-O(N)) \le p_{\theta_0}(u) \le 1 - \exp(-O(N)) \; .
\end{align*}

We inspect the construction to understand which private signal configurations
are consistent with the observation history of agent $\OBS$. First,
the signals of all auxiliary agents in the gadgets
can be directly inferred by $\OBS$.
With that in mind, fix a sequence
of private signals to variable agents $S(x_1), \ldots, S(x_n)$.
Abusing notation, we identify such sequence with an assignment
$x_1, \ldots, x_n$ in a natural way.
Variable gadgets ensure 
that each ``negation agent'' received the opposite signal
$S(\lnot x_i) = 1-S(x_i)$. Moreover, due to clause and not-equal gadgets
we have the following:
\begin{claim}
  \label{cl:extension}\mbox{}
  
  \begin{itemize}
  \item For every assignment $x = (x_1, \ldots, x_n)$, there exists
    exactly one consistent configuration of private signals
    with $S(\EVAL) = 1, S(A) = 0, S(B) = 1$.
  \item For every \emph{satisfying} assignment $x$, there exists
    exactly one consistent configuration of private signals with
    $S(\EVAL) = 0, S(A) = 1, S(B) = 0$.
  \item There are no other consistent configurations.
  \end{itemize}
\end{claim}
As a next step, we compare the likelihoods
of configurations corresponding to different assignments. To this end,
we let the quantity
$P(x, S_0, \theta_0)$ be the a priori probability that private signals are
in the consistent configuration corresponding to assignment $x$,
$S(\EVAL) = S_0$ and $\theta = \theta_0$ (note that this is a 
different definition than given in \eqref{eq:10}).
Furthermore, we set
$P(x, \theta_0) := P(x, 0, \theta_0) + P(x, 1, \theta_0)$.

By inspecting the construction in a similar way as in Claims~\ref{cl:threshold}
and~\ref{cl:counting}
we observe that, for any assignment $x$:
\begin{align*}
  P(x, 1, \sT) &= q \cdot 0.9 \cdot \alpha_1^b \cdot (1-\alpha_3^b)\; ,\\
  P(x, 1, \sF) &= q \cdot 0.4 \cdot (1-\alpha_2^b) \cdot \alpha_4^b \; 
\end{align*}
for some $q(N, M) > 0$ that does not depend on a specific assignment $x$.
On the other hand, for any satisfying assignment $x$ we additionally have
\begin{align*}
  P(x, 0, \sT) &= q \cdot 0.1 \cdot (1-\alpha_1^b) \cdot \alpha_3^b \; ,\\
  P(x, 0, \sF) &= q \cdot 0.6 \cdot \alpha_2^b \cdot (1-\alpha_4^b) \; .
\end{align*}
Each of those expressions is a product of four terms.
The value $q$ corresponds to
the probabilities of signals in variable agents and auxiliary agents in
the gadgets.
The other terms arise from private signals of,
respectively, $\EVAL$, $A$ and $B$.

We choose $\alpha_3 := 0.9$, $\alpha_2 := \alpha_4 := 0.6$, $\alpha_1 := 0.4$
and note that our choice of $b = CN$ for large enough $C$
ensures that we can estimate\footnote{
  The bounds below are slightly better than needed in order to facilitate
  the proof of Theorem~\ref{thm:pspace}.
}
\begin{align}
  P(x, 1, \sT) \in q \cdot 0.4^b \cdot \left(1\pm\frac{1}{200N}\right)^b
  \; , \label{eq:01}\\
  P(x, 1, \sF) \in q \cdot 0.6^b \cdot \left(1\pm\frac{1}{200N}\right)^b
  \; , \label{eq:02}
\end{align}
and, for satisfying assignments,
\begin{align}
  P(x, 0, \sT) \in q \cdot 0.9^b \cdot \left(1\pm\frac{1}{200N}\right)^b \; . \label{eq:03}\\
  P(x, 0, \sF) \in q \cdot 0.6^b \cdot \left(1\pm\frac{1}{200N}\right)^b \; . \label{eq:04}
\end{align}
This in turn implies that for a satisfying assignment we have
\begin{align}
  P(x, \sT) \in q \cdot 0.9^b \cdot \left(1 \pm \frac{1}{100N}\right)^b \; ,
  P(x, \sF) \in q \cdot 0.6^b \cdot \left(1 \pm \frac{1}{100N}\right)^b \; ,
  \label{eq:13}
\end{align}
and for an unsatisfying one
\begin{align}
  P(x, \sT) \in q \cdot 0.4^b \cdot \left(1 \pm \frac{1}{100N}\right)^b \; ,
  P(x, \sF) \in q \cdot 0.6^b \cdot \left(1 \pm \frac{1}{100N}\right)^b \; .
  \label{eq:14}
\end{align}

Accordingly, if the formula $\phi$ has a satisfying assignment $x^*$,
it must be that the belief of agent $\OBS$ at the significant time $t=2$ can be
bounded by
\begin{align}
  \label{eq:07}
  1-\mu(\OBS)
  = \frac{\sum_{x \in \{0,1\}^N} P(x, \sF)}{\sum_{x \in \{0,1\}^N} P(x,\sF)+P(x,\sT)}
  \le \frac{\sum_{x\in\{0,1\}^N} P(x, \sF)}{P(x^*, \sT)}
  \le \frac{2^{N}\cdot 0.61^b}{0.89^b} \le 0.69^b \; .
\end{align}
At the same time, this probability can be lower bounded as
\begin{align}
  \label{eq:08}
  1-\mu(\OBS)
  &\ge \frac{P(x^*, \sF)}{\sum_{x\in\{0,1\}^N} P(x, \sT)+P(x, \sF)}
  \ge \frac{0.59^b}{2^{N+1}\cdot 0.91^b} \ge 0.64^b \; .
\end{align}

If the formula $\phi$ is not satisfiable, a simpler computation
taking into account only equation \eqref{eq:14}
gives
\begin{align}
  \label{eq:12}
  \mu(\OBS)
  \in [0.64^b, 0.69^b]\; .
\end{align}
Hence, $\mu(\OBS) = 1 - \exp(-\Theta(N))$ if $\phi$ is satisfiable
and $\mu(\OBS) = \exp(-\Theta(N))$ otherwise.
\qed

\begin{remark}\label{rem:ties}
  There are some results and proofs about opinion exchange models
  that are sensitive to the tie-breaking rule chosen (see, e.g.,
  Example~3.46 in~\cite{MT17}). We claim that the reduction described above
  (as well as other reductions in this paper) does not suffer
  from this problem.
  
  Ideally, we would like to say that ties never arise in signal configurations
  that are consistent with inputs to the reduction. This is seen to be
  true by inspection, with the following exception:
  Agents that do not receive private signals are indifferent about the
  state of the world until their significant time. We made this choice to
  simplify the exposition. Since significant times are common knowledge,
  no agent places any weight on others' actions before their significant time
  (regardless of the tie-breaking rule used), and the analysis of the reduction
  is not affected in any way by this fact.

  That being said, the ties could be avoided altogether. For example,
  we could introduce an agent $\EPS$ that is observed by everyone else
  at time $t=0$, indicating the action $A(\EPS) = \sT$ and private signal
  $S(\EPS) = 1$ corresponding to
  the likelihood $l_{1}(\EPS) = \eps$ for a small constant $\eps > 0$.
  Since likelihoods arising in the analysis of our reduction are always bounded
  away from zero, $\eps$ can be made small enough so that the agent $\EPS$
  does not affect other agents' actions at their significant times.
  This almost takes care of the problem, except for the agents without private
  signals at time $t=0$ (since they will acquire information from $\EPS$
  only at time $t=1$). This can be solved by giving each such agent $u$
  an informative private signal with likelihoods, say
  \begin{align*}
    \ell_1(u) =
    -\ell_0(u) = \frac{\eps}{100|V|} \; .
  \end{align*}
  In that case $u$ will output an action corresponding to its private signal
  at time $t=0$, but its belief due to private signal
  (and signals of all other non-informative agents that $u$ observes) 
  will become dominated by
  belief of $\EPS$ at time $t=1$.
\end{remark}

\section{\texorpdfstring{$\PSPACE$}{PSPACE}-hardness:
  Proof of Theorem~\ref{thm:pspace}}
\label{sec:pspace}

\paragraph{$\TQBF$ and the high-level idea}
Recall that we will
show $\PSPACE$-hardness by reduction from the
canonical $\PSPACE$-complete language $\TQBF$. More precisely,
we use a representation of quantified Boolean formulas
\begin{align*}
  \Phi = Q_K x_K \cdots Q_1 x_1: \phi(x_K, \ldots, x_1) \; ,
\end{align*}
where:
\begin{itemize}
\item $Q_i$ is a quantifier such that $Q_i \in \{\exists, \forall\}$,
  $Q_i \ne Q_{i+1}$
  and $Q_1 = \exists$.
\item $x_K, \ldots, x_1$ are \emph{blocks of variables} such that their
  total count is $|x_K| + \ldots + |x_1| = N$.
\item $\phi$ is a propositional logical formula given in the 3-CNF form
  with $M$ clauses.
\end{itemize}
The language $\TQBF$ consists of all formulas $\Phi$ that are true.
It is common and useful to think of $\Phi$ as defining a ``position''
in a game, where ``Player 1'' chooses values of variables
under existential quantifiers, ``Player 0'' chooses values of variables
under universal quantifiers, and the objective of Player $s$ is to
evaluate $\phi$ to the value $s$. Under that interpretation,
$\Phi \in \TQBF$ if and only if Player 1 has a winning strategy in
the given position.

Keeping that in mind, we can give an intuition for the proof:
In the $\3SAT$ reduction, if the formula had a satisfying assignment,
then agent $\OBS$ could conclude whp.~that the ``hidden'' assignment
is satisfying, and $\theta=\sT$. Otherwise, the hidden assignment is
not satisfying and $\theta=\sF$ whp. In the $\PSPACE$ reduction,
the hidden assignment will correspond (whp.) to a ``transcript'' of the
game played according to a winning strategy for one of
the players, and $\theta$ will be determined by the winning player.
This will be achieved by implementing a sequence of observer
agents $\OBS_1, \ldots, \OBS_K$, where:
\begin{itemize}
\item Ultimately, the hardness will be shown for the computation of agent
  $\OBS_K$.
\item Agent $\OBS_i$ directly observes variable agents in
  blocks $x_K, \ldots, x_{i+1}$.
\item For each $i$, there is a (slightly more complicated) gadget
  similar to the ``$(\EVAL, A, B)$-gadget'' employed in the $\3SAT$ reduction.
  This gadget involves $\OBS_{i-1}$ as well as two new agents $A_i$ and $B_i$
  and is observed by $\OBS_i$.
  Its purpose is to ``flip'' relative
  likelihoods of different types of variable assignments
  to implement a quantifier switch.
\end{itemize}

\paragraph{The reduction} Recall our formula
\begin{align*}
  \Phi = Q_Kx_K \cdots \exists x_1: \phi(x_K, \ldots, x_1) \; .
\end{align*}
The reduction is defined inductively, with the overall structure illustrated
in Figure~\ref{fig:pspace-reduction}. First, we make a network
identical to the one used used in $\3SAT$ reduction for the formula
$\phi(x_K, \ldots, x_1)$ (i.e., as if all variables were existential).
We call the observer agent $\OBS_1$ and introduce one difference:
$\OBS_1$ additionally directly observes all variable agents
in variable blocks $x_K, \ldots, x_2$.

Next, for each $1 < i \le k$ we place two 
agents $A_i$ and $B_i$ with private signals according to probabilities
$p_\sT(A_i) := 1-\alpha_1^b$, $p_\sF(A_i) := \alpha_2^b$,
$p_\sT(B_i) := 1-\alpha_3^b$, $p_\sF(B_i) := \alpha_4^b$.
The parameter $b$ is the same as in the $\3SAT$ reduction, i.e.,
$b = C \cdot N$ for some absolute $C$ big enough.
The $\alpha_j$ values depend on the parity of $i$ and are provided in
Table~\ref{tab:alpha}.
\begin{table}[ht]\centering
  \caption{Values of $\alpha_j$ for even and odd $i$.}
  \label{tab:alpha}
  \renewcommand{\arraystretch}{1.2}
  \begin{tabular}{|r|l|l|}
    \hline
    &even $i$&odd $i$\\ \hline
    $\alpha_1$&$\frac{4}{9} \cdot 0.9$&$0.9$\\ \hline
    $\alpha_2$&$0.9$&$\frac{4}{9}\cdot 0.9$\\ \hline
    $\alpha_3$&$0.9$&$\frac{4}{9} \cdot 0.9$\\ \hline
    $\alpha_4$&$\frac{4}{9} \cdot 0.9$&$0.9$\\ \hline
    $\delta$ for ``large'' threshold&$0.2b$&$b$\\ \hline
    $\delta$ for ``small'' threshold&$-b$&$-0.2b$\\ \hline
  \end{tabular}
\end{table}

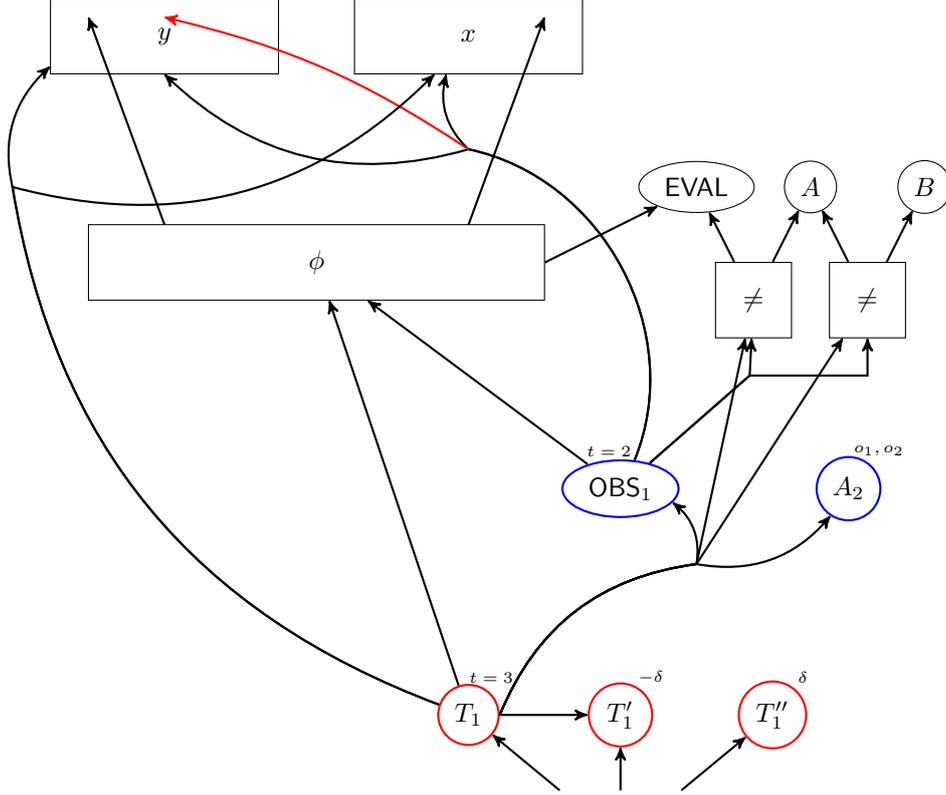
\begin{figure}[ht]\centering
  \caption[LoF entry]{Modified threshold agent illustrated on case $i=K=2$
    and formula $\Phi = \forall y \exists x: \phi(y, x) = 1$.
    The gadget consists of nodes $T_1$, $T_1'$ and $T_1''$.
    These three agents serve the role of $A$, $B$ and $C$ from
    Figure~\ref{fig:threshold-gadget} and are all observed
    by agent $\OBS_{2}$
    (see Figure~\ref{fig:pspace-reduction}). The gadget implements
    ``not-equal'' behavior between agents $\OBS_1$ and $A_2$.
    
    Red arrow emphasizes that agent $\OBS_1$ directly observes
    variable agents associated with $y$.
    Some significant times and likelihoods are shown.
  }
  \label{fig:ad-hoc-threshold}
  \begin{tikzpicture}
    \node [draw, minimum width=3cm, minimum height=1cm]
    (x) at (0, 10) {$y$};
    \node [draw,  minimum width=3cm, minimum height=1cm]
    (y) at (4, 10) {$x$};
    \node [draw, minimum width=6cm, minimum height=1cm]
    (phi) at (2, 7) {$\phi$};
    \node [draw, ellipse] (EVAL) at (7, 8) {$\EVAL$};
    \node [draw, circle] (A) at (8.5, 8) {$A$};
    \node [draw, circle] (B) at (10, 8) {$B$};
    \node [draw, minimum width=1cm, minimum height=1cm]
    (NE1) at (7.75, 6.5) {$\ne$};
    \node [draw, minimum width=1cm, minimum height=1cm]
    (NE2) at (9.25, 6.5) {$\ne$};
    \node [draw=blue, thick, ellipse] (OBS1) at (6, 4) {$\OBS_1$};
    \node [font=\fontsize{6}{0}\selectfont, align=center]
    at (5.85, 4.5) {$t=2$};
    \node [draw=blue, thick, circle] (A1) at (9, 4) {$A_2$};
    \node [font=\fontsize{6}{0}\selectfont, align=center]
    at (9.4, 4.5) {$o_1,o_2$};
    \node [draw=red, thick, circle] (T1) at (4, 1) {$T_1$};
    \node [font=\fontsize{6}{0}\selectfont, align=center]
    at (4.3, 1.5) {$t=3$};
    \node [draw=red, thick, circle, draw] (T') at (6, 1) {$T'_1$};
    \node [font=\fontsize{6}{0}\selectfont, align=center]
    at (6.4, 1.5) {$-\delta$};
    \node [draw=red, thick, circle] (T'') at (8, 1) {$T''_1$};
    \node [font=\fontsize{6}{0}\selectfont, align=center]
    at (8.4, 1.5) {$\delta$};

    \draw [myArrow] (0, 7.5) to (-1, 10.25);
    \draw [myArrow] (4, 7.5) to (5, 10.25);
    \draw [myArrow] (phi.east) to (EVAL);
    \draw [myArrow] (NE1) to (EVAL);
    \draw [myArrow] (NE1) to (A);
    \draw [myArrow] (NE2) to (A);
    \draw [myArrow] (NE2) to (B);
    \draw [myArrow] (OBS1) to (7.7, 5.5) to (NE1);
    \draw [myArrow] (OBS1) to (7.7, 5.5) -| (NE2);
    \draw [myArrow] (OBS1) to (phi);
    \draw [myArrow] (OBS1) to [bend right=50] (4, 8.5)
    to [bend left] (y);
    \draw [myArrow] (OBS1) to [bend right=50] (4, 8.5)
    to [bend left] (x.south);
    \draw [myArrow, color=red] (4, 8.5) to [bend right=10] (0, 10.25);
    \draw [myArrow] (T1) to (T');
    \draw [myArrow] (T1.east) to [bend left] (7, 3) to [bend right] (OBS1);
    \draw [myArrow] (T1.east) to [bend left] (7, 3) to [bend right] (A1);
    \draw [myArrow] (T1.east) to [bend left] (7, 3) to (NE1);
    \draw [myArrow] (T1.east) to [bend left] (7, 3) to (NE2);
    \draw [myArrow] (T1) to (phi);
    \draw [myArrow] (T1) to [bend left] (-2, 8) to [bend left] (x);
    \draw [myArrow] (T1) to [bend left] (-2, 8) to [bend right] (y);
    \draw [myArrow] (5.2, 0) to (T1);
    \draw [myArrow] (6, 0) to (T');
    \draw [myArrow] (6.8, 0) to (T'');
  \end{tikzpicture}
\end{figure}

\begin{figure}[ht]\centering
  \caption[LoF entry]{Schematic representation of the network
    in case $K=2$ for formula $\Phi = \forall y\exists x:\phi(y, x) = 1$.
    The agents and gadgets added in the inductive definition for $i=2$ are
    marked in blue. For clarity, edges from the modified not-equal gadget
    (cf.~Figure~\ref{fig:ad-hoc-threshold}, here marked with exclamation point)
    are not shown.
  }
  \label{fig:pspace-reduction}
    \begin{tikzpicture}
    \node [draw, minimum width=3cm, minimum height=1cm]
    (x) at (0, 10) {$y$};
    \node [draw,  minimum width=3cm, minimum height=1cm]
    (y) at (4, 10) {$x$};
    \node [draw, minimum width=6cm, minimum height=1cm]
    (phi) at (2, 7) {$\phi$};
    \node [draw, ellipse] (EVAL) at (7, 8) {$\EVAL$};
    \node [font=\fontsize{6}{0}\selectfont] at (7.3, 8.5) {$t=0$};
    \node [draw, circle] (A) at (8.5, 8) {$A$};
    \node [font=\fontsize{6}{0}\selectfont] at (8.8, 8.5) {$t=0$};
    \node [draw, circle] (B) at (10, 8) {$B$};
    \node [font=\fontsize{6}{0}\selectfont] at (10.3, 8.5) {$t=0$};
    \node [draw, minimum width=.6cm, minimum height=.6cm]
    (NE1) at (7.75, 6.5) {$\ne$};
    \node [draw, minimum width=.6cm, minimum height=.6cm]
    (NE2) at (9.25, 6.5) {$\ne$};
    \node [draw, ellipse] (OBS1) at (6, 4) {$\OBS_1$};
    \node [font=\fontsize{6}{0}\selectfont] at (6.1, 4.5) {$t=2$};
    \node [draw=blue, thick, circle] (A1) at (8.5, 4) {$A_2$};
    \node [font=\fontsize{6}{0}\selectfont] at (8.8, 4.6) {$t=2$};
    \node [draw=blue, thick, circle] (B1) at (11, 4) {$B_2$};
    \node [font=\fontsize{6}{0}\selectfont] at (11.3, 4.6) {$t=2$};
    \node [draw=blue, thick, minimum width=.6cm, minimum height=.6cm]
    (NE3) at (7.25, 2.5) {$!\ne$};
    \node [draw=blue, thick, minimum width=.6cm, minimum height=.6cm]
    (NE4) at (9.75, 2.5) {$\ne$};
    \node [draw=blue, thick, ellipse] (OBS2) at (3, 2) {$\OBS_2$};

    \draw [myArrow] (0, 7.5) to (-1, 10.25);
    \draw [myArrow] (4, 7.5) to (5, 10.25);
    \draw [myArrow] (phi.east) to (EVAL);
    \draw [myArrow] (NE1) to (EVAL);
    \draw [myArrow] (NE1) to (A);
    \draw [myArrow] (NE2) to (A);
    \draw [myArrow] (NE2) to (B);
    \draw [myArrow] (OBS1.east) to (7.7, 5.5) to (NE1);
    \draw [myArrow] (7.7, 5.5) -| (NE2);
    \draw [myArrow] (OBS1) to (phi);
    \draw [myArrow] (OBS1.north east) to [bend right=50] (4, 8.5)
    to [bend left] (y);
    \draw [myArrow] (4, 8.5) to [bend left] (x.south);
    \draw [myArrow, color=red] (4, 8.5) to [bend right=10] (0, 10.25);
    \draw [myArrow] (NE3) to (OBS1);
    \draw [myArrow] (NE3) to (A1);
    \draw [myArrow] (NE4) to (A1);
    \draw [myArrow] (NE4) to (B1);
    \draw [myArrow] (OBS2) |- (7.25, 1.5) to (NE3);
    \draw [myArrow] (OBS2) |- (7.25, 1.5) -| (NE4);
    \draw [myArrow] (OBS2) to [bend right] (7.75, 4.6) to (NE1.south east);
    \draw [myArrow] (OBS2) to [bend right] (7.75, 4.6) to (NE2.south west);
    \draw [myArrow] (OBS2) to (phi);
    \draw [myArrow] (OBS2) to [bend left] (-2, 7.5) to [bend left] (x);
    \draw [myArrow] (OBS2) to [bend left] (-2, 7.5) to [bend right=10] (y);
  \end{tikzpicture}
\end{figure}
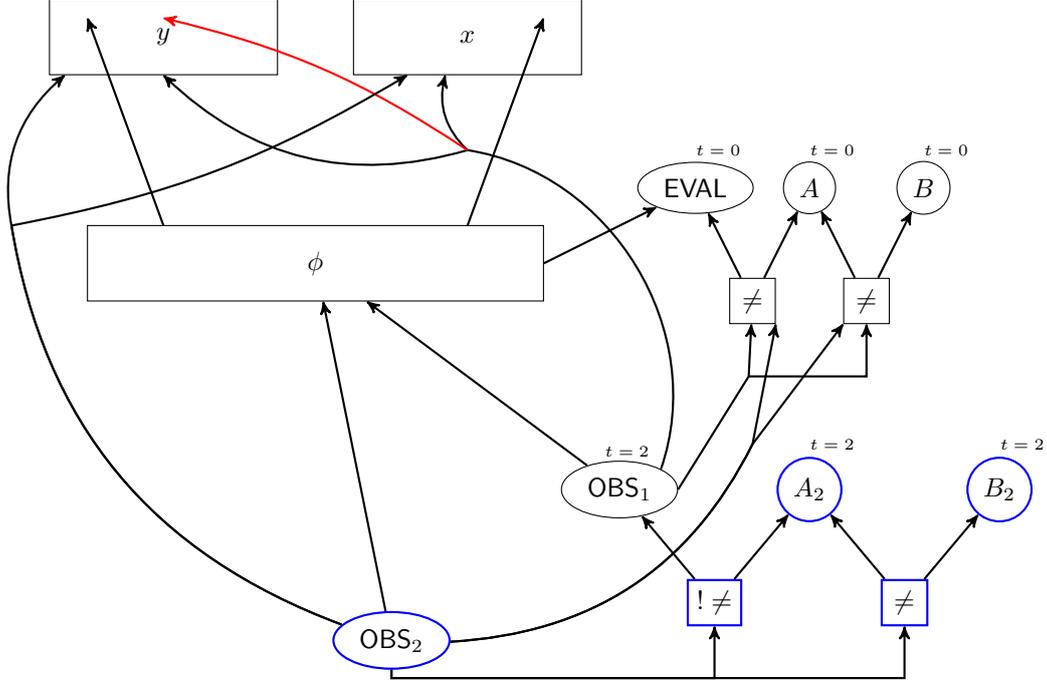

We place a not-equal gadget between $A_i$ and $B_i$. This agent will be observed
by $\OBS_i$.
We would also like to place a not-equal gadget between $\OBS_{i-1}$ and $A_i$.
More precisely, we want a gadget that will reveal that relevant actions
are different: $A(\OBS_{i-1}) \ne A(A_i)$.
We cannot use the standard not-equal gadget directly,
since $\OBS_{i-1}$ receives more complicated
information than a single private signal.
We now describe how to overcome this difficulty, with an illustration 
in Figure~\ref{fig:ad-hoc-threshold}.

We put in place a gadget with a structure analogous to the not-equal gadget
between $\OBS_{i-1}$ and $A_i$.
We will call it a \emph{modified not-equal gadget}.
It consists of two \emph{modified threshold gadgets}. One of those gadgets
ensures that $A(\OBS_{i-1}) \ne \sT$ or $A(A_i) \ne \sT$ (we will call it
a ``large'' threshold), and the other one ensures that
$A(\OBS_{i-1}) \ne \sF$ or $A(A_i) \ne \sF$ (this is a ``small'' threshold).
Of course the conjunction of those two guarantees is equivalent to
$A(\OBS_{i-1}) \ne A(A_i)$.
Since the analysis of two threshold gadgets is symmetric,
we describe only the large threshold.

We call the main, ``summing'' agent of this threshold gadget
$T_i$ (it is an equivalent of $A$ in
Figure~\ref{fig:threshold-gadget}). The agent $T_i$:
\begin{itemize}
\item Observes agents $\OBS_{i-1}$ and $A_{i}$.
\item Additionally observes all agents that $\OBS_{i-1}$ observes.
\item \emph{Except} that it does not observe variable agents in
  variable block $x_i$.
\end{itemize}
The significant time of agent $\OBS_{i-1}$ is $t=2i-2$ and we
set significant time of $T_i$ to $t=2i-1$.

Furthermore, we place two more agents $T'_i$ and $T''_i$
corresponding to agents $B$ and $C$ in Figure~\ref{fig:threshold-gadget}.
Agent $T'_i$ is observed by $\OBS_i$ and $T_i$, and broadcasts
likelihood $-\delta$. Agent $T''_i$ is observed only by $\OBS_i$
and broadcasts likelihood $\delta$.
We still need to define the threshold value $\delta$.
This is not immediate, since we only have bounds
\eqref{eq:07}--\eqref{eq:12} on beliefs of
agent $\OBS_{i-1}$, but it can be done.
Precise values for both large and small thresgholds are in Table~\ref{tab:alpha}.

Finally, we place an agent $\OBS_i$ that observes the same agents
as $\OBS_{i-1}$, except for variable agents in variable block $x_i$.
Note that $\OBS_i$ does not directly observe $\OBS_{i-1}$.
Additionally, $\OBS_i$ observes
the two not-equal gadgets defined above. It does not
receive a private signal, and its significant time is $t=2i$.

This concludes the definition of the reduction.
We show hardness for the computation of agent $\OBS_K$ at time
$t=2K$. Again, since this agent observes only gadgets, its observation history
is naturally determined by the semantics of the gadgets.
We will show that the truth value of formula $\Phi$ reduces to distinguishing
between $\mu(\OBS_K) \approx 1$ and $\mu(\OBS_K) \approx 0$
and, by implication, $A(\OBS_K) = \sT$ and $A(\OBS_K) = \sF$.

\paragraph{Analysis: Preliminaries}
To start with, we note that the $i$-th stage of the inductive definition
adds $O(i(N+M))$ new agents (remembering that there are dummy agents that
are not shown in the figures). Consequently, the total number of agents
is $O(K^2(N+M)) \le O(N^2(N+M))$.
Furthermore, the signal probabilities satisfy \eqref{eq:21} by design.

To analyse the belief of agent $\OBS_K$,
we need to start with more notation and definitions.
For $i > 1$ and
a partial assignment to variable blocks $y := (y_K, \ldots, y_i)$,
let $\Phi_{y}$ be the formula
\begin{align*}
  \Phi_{y} := Q_{i-1}x_{i-1} \cdots \exists x_1:
  \phi(y_k, \ldots, y_i, x_{i-1}, \ldots, x_1)\; ,
\end{align*}
i.e., the original formula with ``hard-coded'' values of $y$.

Let $G_i$ be the part of the network consisting of all agents created up to
the $i$-th step of our inductive definition. Therefore,
$G = G_K \supseteq \ldots \supseteq G_1$.
The network was defined so that all actions of agents in $G_i$
depend only on private signals of agents in $G_i$.
Furthermore, the belief $\mu(\OBS_i)$ depends only
on private signals
of variable agents in $x_K, \ldots, x_{i+1}$
and observations of gadgets by $\OBS_i$ (with the latter
determined by the reduction, since $\OBS_K$ observes all those gadgets as well).

We now need a careful definition in similar vein to $P(x, \theta_0)$ from
the $\3SAT$ reduction.
Given $i$, $1 \le i \le K$, an assignment
$(y, x) := (y_K, \ldots, y_{i+1}, x_i, \ldots, x_1)$, as well as
$\theta_0 \in \{\sT, \sF\}$
we let $P_i(y, x, \theta_0)$ as the probability
that all of the following hold:
\begin{enumerate}
\item For all gadgets observed by the agent $\OBS_i$, $\OBS_i$ observed the
  actions given by the reduction.
\item The assignment determined by private signals of variable agents
  is equal to $(y, x)$.
\item State of the world is $\theta=\theta_0$.
\end{enumerate}
One checks that $P_i(y, x, \theta_0)$ depends only on private signals in $G_i$.
To gain intuition, the reader is invited to convince themselves
that, provided that the modified not-equal gadget ensures
$A(\OBS_{i-1}) \ne A(A_i)$ (we still need to prove that),
$P_i(y, x, \theta_0)$ is always a sum over probabilities of one
(if $\phi(y,x)=0$)
or two (in case $\phi(y,x)=1$) signal
configurations on $G_i$.

Finally, given $y$ and $\alpha, \eps \in (0, 1)$ we will say
that state of the world $\theta_0$ is $\alpha$-likely with error $\eps$ if both
\begin{align*}
  \exists x:
  P_i(y, x, \theta_0) \ge
    \alpha^b \cdot (1 - \eps)^b \; ,\\
  \forall x: P_i(y, x, \theta_0) \le
  \alpha^b \cdot (1 + \eps)^b \; .
\end{align*}

The analysis proceeds by induction on the block number $i$,
with a two-part invariant we need to maintain.
The first part says that,
letting $\eps := \frac{i}{100N}$, there exists some
$\beta := \beta(i) \in (0, 1)$
such that for every partial assignment
$y := (y_K, \ldots, y_{i+1})$:
\begin{enumerate}
\item If $i$ is odd and $\Phi_{y}$ is true, then
  $\sT$ is $\beta$-likely with error $\eps$
  and $\sF$ is $\frac{2}{3}\beta$-likely with error $\eps$.
\item If $i$ is odd and $\Phi_{y}$ is false, then
  $\sT$ is $\frac{4}{9}\beta$-likely with error $\eps$ and
  $\sF$ is $\frac{2}{3}\beta$-likely with error $\eps$.
\item Symmetrically, if $i$ is even and $\Phi_{y}$ is true,
  then $\sT$ is $\frac{2}{3}\beta$-likely with error $\eps$ and $\sF$ is
  $\frac{4}{9}\beta$-likely with error $\eps$.
\item If $i$ is even and $\Phi_{y}$ is false, then
    $\sT$ is $\frac{2}{3}\beta$-likely with error $\eps$
    and $\sF$ is $\beta$-likely with error $\eps$.
\end{enumerate}

The second part of the invariant states that whenever $\Phi_{y}$ is true,
the belief of agent $\OBS_{i}$ satisfies
$1-\mu(\OBS_i) \in [0.64^b, 0.69^b]$.
Similarly, if $\Phi_y$ is false, then this belief
satisfies $\mu(\OBS_i) \in [0.64^b, 0.69^b]$. Note that this part
applied to $i=K$ implies the last bullet point in the statement
of Theorem~\ref{thm:pspace}, with $\mu(\OBS_K)$
being within $\exp(-\Theta(N))$ distance to either zero or one.

\paragraph{Base case}
To establish the base case $i=1$ one has to go through the proof
in Section~\ref{sec:np} and convince themselves that the analysis stays valid
even when the agent $\OBS$ directly observes variable agents
$y_K, \ldots, y_2$. Then, the first invariant
is established with
\begin{align*}
  \beta(1) := q^{1/b} \cdot 0.9 \; ,
\end{align*}
where $q$ is the value featured in equations \eqref{eq:13}-\eqref{eq:14}.
For example, $\Phi_y$ being true means that the respective $3$-CNF formula
$\phi_y(x)$ is satisfiable. Taking a satisfying assignment $x$, we get
by \eqref{eq:13}
\begin{align*}
  P_1(y, x, \sT)
  &= P(x, \sT) \ge q \cdot 0.9^b \cdot (1-\eps)^b
    = \beta^b \cdot (1-\eps)^b \; ,\\
  P_1(y, x, \sF)
  &= P(x, \sF) \ge q \cdot 0.6^b \cdot (1-\eps)^b
    = \left(\frac{2}{3}\beta\right)^b \cdot (1-\eps)^b \;.
\end{align*}
On the other hand, by \eqref{eq:13} and \eqref{eq:14},
for every assignment, satisfying or not, we have
\begin{align*}
  P_1(y, x, \sT)
  &\le \max\left(q\cdot 0.9^b\cdot(1+\eps)^b, q\cdot 0.4^b \cdot (1+\eps)^b
    \right)
    \le \beta^b \cdot (1+\eps)^b \; ,\\
  P_1(y, x, \sF)
  &\le q \cdot 0.6^b \cdot (1+\eps)^b
    = \left(\frac{2}{3}\beta\right)^b \cdot (1+\eps)^b \;.
\end{align*}
Similar computation gives the first invariant in case $\Phi_y$ is false,
this time using only \eqref{eq:14}. The second invariant is a direct
consequence of equations \eqref{eq:07}-\eqref{eq:12}.

\paragraph{Induction step}
We will analyze only even $i$, since the other case is analogous.
Fix some $y = (y_K, \ldots, y_{i+1})$. In the following we assume
that all actions observed in gadgets are as given by the reduction
and that private signals for the initial blocks of variables are given
by $y$. Let us call private signal configurations on $G_{i-1}$
that satisfy those conditions \emph{consistent}.

In this setting, every assignment of private signals in the block $y_i$
determines the action $A(\OBS_{i-1})$ and, by the second invariant,
$A(\OBS_{i-1}) = \sT$ if and only if $\Phi_{y,y_i}$ is true.
Accordingly, we divide consistent configurations
into ``$\sT$-configurations'' and ``$\sF$-configurations''.

Our first objective is to show that the modified not-equal gadget
(cf.~Figure~\ref{fig:pspace-reduction}) ensures that
$A(\OBS_{i-1}) \ne A(A_i)$.
Let $T_i$ be the main agent in a modified threshold gadget between
$\OBS_{i-1}$ and $A_i$ (cf.~Figure~\ref{fig:ad-hoc-threshold}).
At its significant time $t=2i-1$, agent $T_i$ observed
everything that agent $\OBS_{i-1}$
observed except for the assignment $y_i$. It also observed the action
$A(\OBS_{i-1}) = \theta_0$.
Therefore,  the private signal configurations on $G_{i-1}$
consistent with observations of $T_i$ are exactly the 
$\theta_0$-configurations. We let
\begin{align*}
  p_{\OBS}(\theta_0) := \EE \left[ \mu(\OBS_{i-1}) \right] \; ,
\end{align*}
where the expectation is over all $\theta_0$-configurations. By the second
invariant, 
$p_{\OBS}(\theta_0)$ is at a distance between $0.64^b$ and $0.69^b$
to $1$ or $0$, depending on the value of $\theta_0$.
Let $m(\theta_0):= \ln \frac{p_{\OBS}(\theta_0)}{1-p_{\OBS}(\theta_0)}$.
We check that
\begin{align}
  \label{eq:18}
  m(\sT) \in [0.37b, 0.45b], \; m(\sF) \in [-0.45b, -0.37b] \; .
\end{align}

Recall that, outside of $G_{i-1}$, agent $T_i$ observes
actions (and infers private signals)
of agents $A_i$ and $T'_i$. The likelihoods of $A_i$
are given by
\begin{align*}
  \ell_\sT(A_i)
  &= \ln \frac{1-\alpha_1^b}{\alpha_2^b}
    =\ln\frac{1-\left(\frac{4}{9}\cdot0.9\right)^b}{0.9^b}
    \in [0.1b, 0.11b]\\
  \ell_\sF(A_i)
  &= \ln\frac{\alpha_1^b}{1-\alpha_2^b}
    =\ln\frac{\left(\frac{4}{9}\cdot0.9\right)^b}{1-0.9^b}
    \in [-0.92b, -0.91b]\;.
\end{align*}
The likelihood $-\delta$ of $T'_i$ is given by Table~\ref{tab:alpha}.
Let $\theta_1 := A(\OBS_{i-1})$ and $\theta_2 := A(A_i)$.
Since private signals in $G_{i-1}$,
$A_i$ and $T'_i$ are independent, it must be that
$A(T_i) = \sT$ if and only if
\begin{align}
  \label{eq:16}
  m(\theta_1) + \ell_{\theta_2}(A_i) - \delta > 0 \; .
\end{align}
A calculation shows that the values in Table~\ref{tab:alpha} ensure
$A(OBS_{i-1}) \ne A(A_i)$. For example, for the ``large'' threshold
we have $\delta = 0.2b$, so $\theta_1 = \theta_2 = \sT$ gives
\begin{align*}
  m(\sT) + \ell_{\sT}(A_i) - \delta > (0.37 + 0.1 - 0.2)\cdot b > 0 \; .
\end{align*}
On the other hand, if $\theta_1 \ne \sT$ or $\theta_2 \ne \sT$, then
\begin{align*}
  m(\theta_1) + \ell_{\theta_2}(A_i) - \delta < (-0.26 - 0.2)\cdot b < 0 \; .
\end{align*}
Performing a similar reasoning for the ``small'' threshold, we can conclude
that in every consistent configuration
$A(\OBS_{i-1}) \ne A(A_i)$, as well as $A(A_i) \ne A(B_i)$.
Therefore, every consistent signal configuration on $G_{i-1}$
can be extended to a unique configuration on $G_i$ that is consistent with observations
of $\OBS_i$.
Consulting Table~\ref{tab:alpha} again,
we compute:
\begin{align}
  \Phi_{y,y_i} \in \TQBF \implies P_i(y, y_i, x; \sT)
  &= P_{i-1}(y, y_i, x; \sT) \cdot q \cdot \alpha_1^b \cdot (1-\alpha_3^b)\label{eq:05}\\
  &\in P_{i-1} (y, y_i, x; \sT) \cdot q \cdot \left(0.9\cdot \frac{4}{9}\right)^b
    \cdot \left(1\pm\frac{1}{200N}\right)^b\nonumber\\
  P_i(y, y_i, x; \sF)
  &\in P_{i-1} (y, y_i, x; \sF) \cdot q \cdot \left(0.9\cdot \frac{4}{9}\right)^b
    \cdot \left(1\pm\frac{1}{200N}\right)^b\nonumber\\
  \Phi_{y,y_i} \notin \TQBF \implies P_i(y, y_i, x; \theta_0)
  &\in P_{i-1} (y, y_i, x; \theta_0) \cdot q \cdot 0.9^b
    \cdot \left(1\pm\frac{1}{200N}\right)^b\label{eq:06}\;,
\end{align}
where $q$ is a universal factor coming from private signals in auxiliary
agents in the not-equal gadgets.

Recall that the first invariant tells us that for
some $\beta' = \beta(i-1) \in (0, 1)$,
if $\Phi_{y, y_i}$ is true, then $\sT$ is $\beta'$-likely
and $\sF$ is $\frac{2}{3}\beta'$-likely, and if $\Phi_{y, y_i}$ is false,
then $\sT$ is $\frac{4}{9}\beta'$-likely and
$\sF$ is $\frac{2}{3}\beta'$-likely, all with error $\frac{i-1}{100N}$.

Note that $\Phi_y$ is true if and only if $\Phi_{y, y_i}$ is true for all $y_i$.
Equivalently, $\Phi_y$ is false if and only if there exists $y_i$ such that
$\Phi_{y, y_i}$ is false. Take
$\beta := \beta(i) := \beta' \cdot q^{1/b} \cdot 0.9 \cdot \frac{2}{3}$.
Then, the first invariant implies
that if $\Phi_y$ is true, then $\sT$ is $\frac{2}{3}\beta$-likely and
$\sF$ is $\frac{4}{9}\beta$-likely, with error $\frac{i}{100N}$.
To see that, note that, by \eqref{eq:06} and induction, we have
that for every $y, y_i, x$:
\begin{align*}
  P_i(y,y_i,x,\sT)
  &\le P_{i-1}(y,y_i,x,\sT)\cdot q\cdot
    \left(0.9\cdot\frac{4}{9}\right)^b
    \left(1+\frac{1}{200N}\right)^b\\
  &\le \left(\beta'\right)^b
    \left(1+\frac{i-1}{100N}\right)^b \cdot q \cdot
    \left(0.9\cdot\frac{4}{9}\right)^b\left(1+\frac{1}{200N}\right)^b\\
  &\le\left(\frac{2}{3}\cdot\beta\right)^b(1+\eps)^b\;,\\
  P_i(y,y_i,x,\sF)
  &\le \left(\frac{2}{3}\cdot\beta'\right)^b
    \left(1+\frac{i-1}{100N}\right)^b \cdot q \cdot
    \left(0.9\cdot\frac{4}{9}\right)^b\left(1+\frac{1}{200N}\right)^b\\
  &\le\left(\frac{4}{9}\cdot\beta\right)^b(1+\eps)^b\;.
\end{align*}
At the same time, symmetric computations give also that for
every $y$ there exist $y_i$ and $x$
(just take arbitrary $y_i$ and $x$ that exists for $\Phi_{y,y_i}$ by
the first invariant) such that
\begin{align*}
  P_i(y, y_i, x, \sT) \ge \left(\frac{2}{3}\cdot\beta\right)^b (1-\eps)^b \;,\\
  P_i(y, y_i, x, \sF) \ge \left(\frac{4}{9}\cdot\beta\right)^b (1-\eps)^b \;.\\
\end{align*}

On the other hand,
if $\Phi_y$ is false, then we need to consider both \eqref{eq:05} and
\eqref{eq:06} to conclude that
$\sT$ is $\frac{2}{3}\beta$-likely and $\sF$ is $\beta$-likely
with error $\eps$. However, the computation is very similar to the previous
ones and we skip it.
Therefore, we implemented the quantifier switch
and reestablished the first induction invariant.

Finally, we need to use a computation similar as in \eqref{eq:07} and
\eqref{eq:08} to check the second invariant.
If $\Phi_y$ is true, then,
since $\sT$ is $\frac{2}{3}\beta$-likely
and $\sF$ is $\frac{4}{9}\beta$-likely with error $\frac{i}{100N}$,
\begin{align*}
  1-\mu(\OBS_i)
  \le \frac{2^N \beta^b (4/9)^b (1+i/100N)^b}
  {\beta^b (2/3)^b (1-i/100N)^b}
  \le \frac{2^N(2/3)^b1.01^b}{0.99^b}
  \le (2/3)^b \cdot 1.03^b \cdot 2^N \le 0.69^b \; ,\\
  1-\mu(\OBS_i)
  \ge \frac{\beta^b (4/9)^b (1-i/100N)^b}
  {2^N \beta^b (2/3)^b (1+i/100N)^b}
  \ge \frac{2^N(2/3)^b0.99^b}{1.01^b}
  \ge (2/3)^b \cdot 0.97^b \cdot 2^N \ge 0.64^b \; .
\end{align*}
A symmetric computation confirms that the second invariant is preserved
also when $\Phi_y$ is false.\qed

\section{Bounded signals: Proof of Theorem~\ref{thm:bounded}}
\label{sec:bounded}
One could object that our reduction uses private signal distributions
with probabilities that are exponentially close to zero and one.
Given that it is a worst-case reduction, with relevant configurations
arising with exponentially small probability, we do not think this is a
significant issue. In any case, in this section we explain
how to modify the proof of Theorem~\ref{thm:pspace} so that it uses
only a fixed collection of (say, at most fifty) private signal distributions.

Note that the only agents we need to replace are $A, B$ from the
$\3SAT$ reduction, and $A_i, B_i$ from the induction step in the $\PSPACE$
reduction, as well as their associated not-equal gadgets.
We sketch the modifications on one example, since other cases are analogous.
To this end, take even $i$ and consider
$A_i$, $B_i$ and their not-equal gadgets (cf.~Figures~\ref{fig:ad-hoc-threshold}
and~\ref{fig:pspace-reduction}).

Going back to the proof of Theorem~\ref{thm:pspace}, in particular
equations \eqref{eq:05}-\eqref{eq:06},
what we would like to have
is that for every consistent configuration on $G_{i-1}$, there should be
a unique way of extending it to a consistent configuration on $G_i$
such that for an assignment $(y, x)$ and $\theta_0 \in \{\sF, \sT\}$,
\begin{align*}
  P_i(y, x, \theta_0) = P_{i-1}(y, x, \theta_0) \cdot r \cdot
  \begin{cases}
    \alpha_1^b&\text{if $A(\OBS_{i-1}) = \sT$ and $\theta_0 = \sT$,}\\
    \alpha_4^b&\text{if $A(\OBS_{i-1}) = \sT$ and $\theta_0 = \sF$,}\\
    \alpha_3^b&\text{if $A(\OBS_{i-1}) = \sF$ and $\theta_0 = \sT$,}\\
    \alpha_2^b&\text{if $A(\OBS_{i-1}) = \sF$ and $\theta_0 = \sF$,}
  \end{cases}
\end{align*}
for some $r \in (0, 1)$ independent of $(y, x, \theta_0)$. We are going
to achieve this using two independent gadgets corresponding to $A_i$ and
$B_i$.
Again, we only sketch the construction for $A_i$.
What we need, then, is to create a gadget
that extends every consistent configuration on $G_{i-1}$ to a unique
consistent configuration on $G_i$ such that
\begin{align}
  \label{eq:15}
  P_i(y, x, \theta_0) = P_{i-1}(y, x, \theta_0) \cdot r \cdot
  \begin{cases}
    \alpha_1^b&\text{if $A(\OBS_{i-1}) = \sT = \theta_0$,}\\
    \alpha_2^b&\text{if $A(\OBS_{i-1}) = \sF = \theta_0$,}\\
    1&\text{otherwise.}
  \end{cases}
\end{align}

\begin{figure}[ht]\centering
  \caption{Bounded signals gadget. One out of $b$ parts is shown. The details
    of the modified not-equal gadget between $\OBS_{i-1}$ and $F$
    are not shown, and the counting gadget between $F$ and $F'$ is not included.
  }
  \label{fig:bounded-gadget}
  \begin{tikzpicture}
    \node [draw, ellipse] (OBSi) at (0, 10) {$\OBS_{i-1}$};
    \node [draw, circle, label=center:$F$] (B) at (2, 10) {\phantom{$C_j$}};
    \node [font=\fontsize{6}{0}\selectfont] at (2.3, 10.6) {$3/4, 1/4$};
    \node [draw, circle] (C) at (5, 10) {$C_j$};
    \node [font=\fontsize{6}{0}\selectfont] at (5.3, 10.6) {$q_\sT, q_\sF$};
    \node [draw, circle] (D) at (7, 10) {$D_j$};
    \node [font=\fontsize{6}{0}\selectfont] at (7.3, 10.6) {$p_\sT, p_\sF$};
    \node [draw, circle] (E) at (9, 10) {$E_j$};
    \node [font=\fontsize{6}{0}\selectfont] at (9.3, 10.6) {$q_\sT, q_\sF$};
    \node [font=\fontsize{15}{0}\selectfont] at (3.2, 9.3) {$\vdots$};
    \node [font=\fontsize{15}{0}\selectfont] at (3.2, 8.35) {$\vdots$};
    \node [font=\fontsize{15}{0}\selectfont] at (3.2, 7.4) {$\vdots$};
    \node [draw, ellipse] (OBS) at (4.5, 6.5) {$\OBS_i$};
    
    \node [draw, minimum width=.6cm, minimum height=.6cm] (NE1)
    at (4, 8.5) {$\ne$};
    \node [draw, minimum width=.6cm, minimum height=.6cm] (NE2)
    at (6, 8.5) {$\ne$};
    \node [draw, minimum width=.6cm, minimum height=.6cm] (NE3)
    at (8, 8.5) {$\ne$};
    \node [draw=blue, thick, minimum width=.6cm, minimum height=.6cm]
    (NES) at (1, 8.5) {$!\ne$};

    \draw [thick] (NE1) to (3.4, 8.95);
    \draw [myArrow] (3, 9.25) to (B);
    \draw [myArrow] (NE1) to (C);
    \draw [myArrow] (NE2) to (C);
    \draw [myArrow] (NE2) to (D);
    \draw [myArrow] (NE3) to (D);
    \draw [myArrow] (NE3) to (E);
    \draw [myArrow] (NES) to (OBSi);
    \draw [myArrow] (NES) to (B);
    \draw [myArrow] (3, 7.37) to (NES);
    \draw [thick] (OBS) to (3.4, 7.12);
    \draw [myArrow] (OBS) to (NE1);
    \draw [myArrow] (OBS) to (NE2);
    \draw [myArrow] (OBS) to (NE3);
  \end{tikzpicture}
\end{figure}
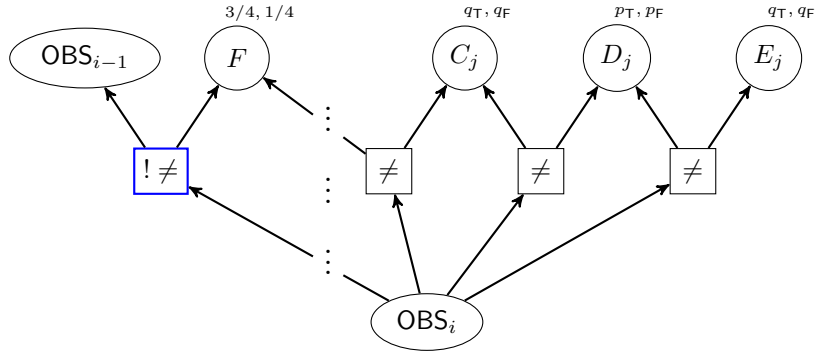

This is achieved as shown in Figure~\ref{fig:bounded-gadget}. We
create an agent $F$ with fixed, arbitrary distribution,
say $p_\sF(F) = 1/4$ and $p_\sT(F) = 3/4$. Then, we add agents
$C_j, D_j, E_j$ for $j = 1,\ldots,b$
with private signal distributions
\begin{align*}
  p_{\theta_0}(D_j) := p_{\theta_0} \; ,\\
  p_{\theta_0}(C_j) := p_{\theta_0}(E_j) := q_{\theta_0} \; ,  
\end{align*}
for some ($\alpha_i$-dependent) constants $p_{\sF}, p_{\sT}, q_{\sF}, q_{\sT}$
that we will specify shortly.

For each triple $C_j, D_j, E_j$ we also place three not-equal gadgets
observed by $\OBS_i$: Respectively, between $F$ and $C_j$,
$C_j$ and $D_j$, and $D_j$ and $E_j$. We also create an agent $F'$ with
the same signal distribution as $F$, and a counting gadget with equivalence
observed by $\OBS_i$, making sure that $S(F)+S(F') = 1$
(this is to get rid of a small distortion
in \eqref{eq:15} due to signal of agent $F$; we will not worry about
it from now on). Finally, we place a gadget between $\OBS_{i-1}$ and $F$
generalizing the modified not-equal gadget from Theorem~\ref{thm:pspace}.
This gadget will be observed by $\OBS_i$ and we will fill in its details later.

Let us assume for now that the modified not-equal gadget
ensures that $A(\OBS_{i-1}) \ne A(F)$ in every consistent configuration.
Then, since not-equal gadgets guarantee $S(F) = S(D_j) \ne S(C_j) = S(E_j)$
for every $j$, we claim that it is not difficult to see that
every consistent configuration on $G_{i-1}$ can be uniquely extended to
consistent configuration on $G_i$ such that
\begin{align*}
  P_i(y, x, \theta_0) = P_{i-1}(y, x, \theta_0) \cdot r \cdot
  \begin{cases}
    \left(q_\sT^2 (1-p_\sT) \right)^b
      &\text{if $A(\OBS_{i-1}) = \sT$ and $\theta_0 = \sT$,}\\
    \left(q_\sF^2 (1-p_\sF) \right)^b
      &\text{if $A(\OBS_{i-1}) = \sT$ and $\theta_0 = \sF$,}\\
    \left((1-q_\sT)^2 p_\sT \right)^b
      &\text{if $A(\OBS_{i-1}) = \sF$ and $\theta_0 = \sT$,}\\
    \left((1-q_\sF)^2 p_\sF \right)^b
      &\text{if $A(\OBS_{i-1}) = \sF$ and $\theta_0 = \sF$.}
  \end{cases}
\end{align*}

Comparing with \eqref{eq:15}, we need to find $p_\sF, p_\sT, q_\sF, q_\sT$
satisfying
\begin{align}
  \label{eq:17}
  \frac{q_\sT^2(1-p_\sT)}{\alpha_1} = q_\sF^2(1-p_\sF) = (1-q_\sT)^2p_\sT
  = \frac{(1-q_\sF)^2p_\sF}{\alpha_2}\; .
\end{align}
Separately comparing and transforming the terms in \eqref{eq:17}:
second with fourth, and then first with third, we get
\begin{align*}
  p_\sF = \frac{\alpha_2 q_\sF^2}{\alpha_2 q_\sF^2+(1-q_\sF)^2} \; ,
  \qquad \qquad
  p_\sT = \frac{q_\sT^2}{q_\sT^2 + \alpha_1(1-q_\sT)^2} \; ,
\end{align*}
which can be substituted into comparison of the first and second term,
yielding
\begin{align*}
  \frac{q_\sT^2(1-q_\sT)^2}{q_\sT^2 + \alpha_1(1-q_\sT)^2} =
  \frac{q_\sF^2(1-q_\sF)^2}{\alpha_2q_\sF^2 + (1-q_\sF)^2} \; .
\end{align*}
Taking $q_\sT := 1-\eps$ for small enough $\eps > 0$, this can be checked to have
a solution with $q_\sF = \eps + O(\eps^2)$,
$p_\sF = \alpha_2\eps^2 + O(\eps^3)$ and $p_\sT = 1 - \alpha_1\eps^2 + O(\eps^3)$.

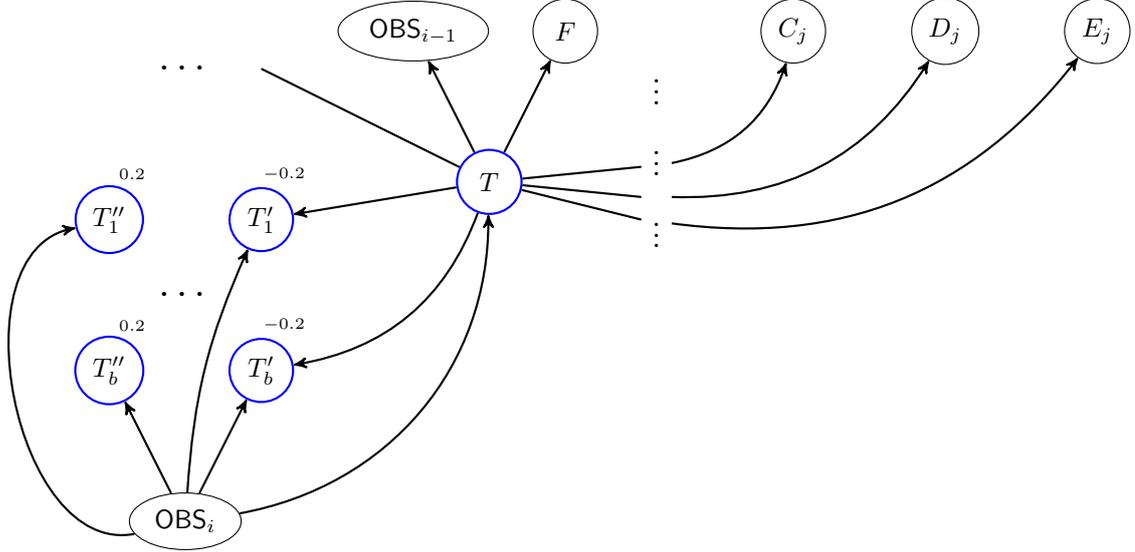
\begin{figure}[ht]\centering
  \caption{Implementation of the modified not-equal gadget (marked in blue
    in Figure~\ref{fig:bounded-gadget}).}
  \label{fig:bounded-ad-hoc}
  \begin{tikzpicture}
    \node [draw, ellipse] (OBSi) at (0, 10) {$\OBS_{i-1}$};
    \node [draw, circle, label=center:$F$] (B) at (2, 10) {\phantom{$C_j$}};
    \node [draw, circle] (C) at (5, 10) {$C_j$};
    \node [draw, circle] (D) at (7, 10) {$D_j$};
    \node [draw, circle] (E) at (9, 10) {$E_j$};
    \node [font=\fontsize{15}{0}\selectfont] at (3.2, 9.3) {$\vdots$};
    \node [font=\fontsize{15}{0}\selectfont] at (3.2, 8.35) {$\vdots$};
    \node [font=\fontsize{15}{0}\selectfont] at (3.2, 7.4) {$\vdots$};
    \node [draw=blue, thick, circle, label=center:$T$]
    (T) at (1, 8) {\phantom{$C_j$}};
    
    \node [font=\fontsize{15}{0}\selectfont] at (-3, 9.5) {$\cdots$};
    \node [draw=blue, circle, thick] (T1) at (-2, 7.5) {$T'_1$};
    \node [font=\fontsize{6}{0}\selectfont] at (-1.7, 8.1) {$-0.2$}; 
    \node [draw=blue, circle, thick] (T'1) at (-4, 7.5) {$T''_1$};
    \node [font=\fontsize{6}{0}\selectfont] at (-3.7, 8.1) {$0.2$}; 
    \node [font=\fontsize{15}{0}\selectfont] at (-3, 6.5) {$\cdots$};
    \node [draw=blue, circle, thick] (Tb) at (-2, 5.5) {$T'_b$};
    \node [font=\fontsize{6}{0}\selectfont] at (-1.7, 6.1) {$-0.2$}; 
    \node [draw=blue, circle, thick] (T'b) at (-4, 5.5) {$T''_b$};
    \node [font=\fontsize{6}{0}\selectfont] at (-3.7, 6.1) {$0.2$}; 
    
    \node [draw, ellipse] (OBS) at (-3, 3.5) {$\OBS_i$};
    
    \draw [myArrow] (T) to (OBSi);
    \draw [myArrow] (T) to (B);
    \draw [thick] (T) to (3, 8.2);
    \draw [myArrow] (3.4, 8.25) to [bend right](C);
    \draw [thick] (T) to (3, 7.8);
    \draw [myArrow] (3.4, 7.8) to [bend right] (D);
    \draw [thick] (T) to (3, 7.5);
    \draw [myArrow] (3.4, 7.45) to [bend right] (E);
    \draw [thick] (T) to (-2, 9.5);
    \draw [myArrow] (T) to (T1);
    \draw [myArrow] (T) to [bend left] (Tb);
    \draw [myArrow] (OBS) to (Tb);
    \draw [myArrow] (OBS) to (T'b);
    \draw [myArrow] (OBS) to [bend left=90] (T'1);
    \draw [myArrow] (OBS) to [bend left=10] (T1);
    \draw [myArrow] (OBS) to [bend right=40] (T);
  \end{tikzpicture}
\end{figure}

We still need to explain how to construct the modified not-equal gadget ensuring
that $A(\OBS_{i-1}) \ne A(F)$. This is a generalization of the construction
in Figure~\ref{fig:ad-hoc-threshold} and is shown in Figure~\ref{fig:bounded-ad-hoc}.

Yet again, it is achieved by combining two threshold gadgets and we
focus on one of them. Recall from Table~\ref{tab:alpha} that this threshold
was set at $\delta = 0.2b$. The objective is to ensure that $A(T) = A(F) = \sT$ if and only
if an inequality like \eqref{eq:16} holds.

The threshold gadget will have a ``counting agent'' $T$ and auxiliary
agents $T'_1, \ldots, T'_b$ and $T''_1, \ldots, T''_b$.
Auxiliary agents receive private signals with likelihoods
$\ell_0(T'_j) := -0.2$ and $\ell_1(T''_j) := 0.2$. Agent $T$
observes $\OBS_{i-1}$, as well as other gadgets and agents
in the network $G_{i-1}$,
in the same way as agent $T_1$ in Figure~\ref{fig:ad-hoc-threshold}.
Additionally,
it directly observes all agents with private signals in the counting gadget between
$F$ and $F'$, as well as all of $C_j, D_j$ and $E_j$. Finally,
it observes $T'_1, \ldots, T'_b$.
Agent $\OBS_i$ observes $T'_1, \ldots, T'_b, T''_1, \ldots, T''_b$ and $T$.
As expected, we specify that at the significant time
$\OBS_i$ observes actions  $A(T'_j) = \sF$, $A(T''_j) = \sT$ and
$A(T) = \sF$.
Note that we do not need to change the significant time of $\OBS_i$.

Assuming that $A(\OBS_{i-1}) = \theta_1$ and $A(F) = \theta_2$,
we use the same reasoning as in \eqref{eq:16} to compute
the likelihood $m(\theta_1)$ that agent $T$ can infer from looking at
$G_{i-1}$, another likelihood $\ell(\theta_2)$ that can be inferred
from looking at $F, C_j, D_j, E_j$ and the likelihood
$-\delta = -0.2b$ arising from looking at $T'_1, \ldots, T'_b$.
The bounds on $m(\theta_1)$ are the same as in \eqref{eq:18},
and as for $\ell(\theta_2)$, from \eqref{eq:17}
we get, as expected $\ell(\sF) = -b \ln \frac{1}{\alpha_1}$
and $\ell(\sT) = b \ln \frac{1}{\alpha_2}$.

Since the
private signals in these three parts of the graph are conditionally
independent, these likelihoods can be added up to ensure
that $A(T) = \sF$ if and only if
\begin{align*}
  m(\theta_1) + \ell(\theta_2) < \delta \; ,
\end{align*}
which implies, the same as in the proof of Theorem~\ref{thm:pspace},
that in a consistent configuration either $\theta_1 = \sF$ or $\theta_2 = \sF$.

As mentioned, other cases proceed in a similar manner. One difference
is that for agents $A$ and $B$ in the base case ($\3SAT$ reduction),
$\EVAL$ is a simple agent with bounded signal (as opposed to $\OBS_{i-1}$).
However, this is only good news for us: We do not need to implement
the modified not-equal gadget, since a simple not-equal gadget
between $B$ and $\EVAL$ suffices.
\qed

\section{\texorpdfstring{$\HASHP$}{\#P}-hardness of revealed beliefs:
Proof of Theorem~\ref{thm:sharp}} 
\label{sec:sharp}

\paragraph{Reduction} Our reduction uses the DAG structure and the concept
of significant time as
explained in Section~\ref{sec:np}. The general idea is as in Theorem~\ref{thm:np},
with some adaptations to the counting setting and revealed beliefs.
We assume that the agents broadcast beliefs in the form of
likelihoods.

\begin{figure}[ht]\centering
  \caption{Revealed belief reduction: Variable gadget.}
  \label{fig:belief-variable}
  \begin{tikzpicture}
    \node [font=\fontsize{6}{0}\selectfont] at (0.3, 0.6) {$\ell_0, \ell_1$};
    \node [draw, circle, label=center:$x_i$] (xi) at (0, 0) {\phantom{$\lnot x_i$}};
    \node [font=\fontsize{6}{0}\selectfont] at (1.7, 0.6) {$\ell_0, \ell_1$};
    \node [draw, circle] (notxi) at (1.4, 0) {$\lnot x_i$};
    \node [draw, circle] (A) at (-2.5, 0) {\phantom{$\lnot x_i$}};
    \node [font=\fontsize{6}{0}\selectfont] at (-2.5, -0.6) {$-\ell_0-\ell_1$};
    \node [draw, circle] (B) at (0.7, -2) {\phantom{$\lnot x_i$}};
    \node [font=\fontsize{6}{0}\selectfont] at (0.7, -2.6) {$\ell_0+\ell_1$};
    
    \node [draw, circle] (OBS) at (-0.9, -4) {$\OBS$};

    \draw [myArrow] (B) -- (xi);
    \draw [myArrow] (B) -- (notxi);
    \draw [myArrow] (OBS) to [bend right] (0.7, -2.8);
    \draw [myArrow] (OBS) to [bend left] (-2.5, -0.8);
  \end{tikzpicture}
\end{figure}
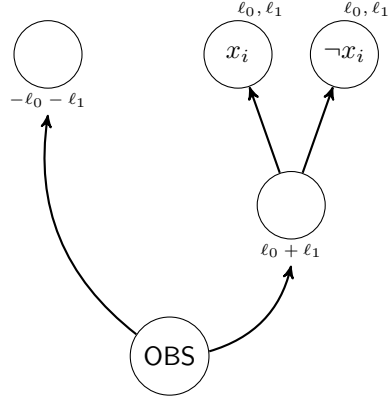

We define a common signal distribution with
$p_\sT := 3/4$ and $p_\sF := 1/4$ and respective
likelihoods $\ell_1$ and $\ell_0$.
The graph we construct contains
an observer agent $\OBS$ with no private signal
and the ``evaluation'' agent $\EVAL$ with $(p_\sT, p_\sF)$ private signal.
Given a $\2SAT$ formula $\phi$ with variables $x_1, \ldots, x_N$
and clauses $C_1, \ldots, C_M$, respective variable and clause gadgets
are designed as follows:

For a variable $x_i$, we create two agents $x_i$ and $\lnot x_i$,
receiving $(p_\sT, p_\sF)$ private signals. Those two agents are observed
by an auxiliary agent, which in turn is observed by agent $\OBS$.
The observation history of $\OBS$ indicates that the auxiliary agent
broadcast likelihood $A = \ell_0 + \ell_1$. At the same time,
$\OBS$ observes another auxiliary agent with informative private signal,
broadcasting likelihood $A = -\ell_0 - \ell_1$.
See Figure~\ref{fig:belief-variable} for illustration. Since the likelihood
broadcast by the agent observing $x_i$ and $\lnot x_i$
is the sum of their likelihoods, we can perform an analysis
similar to the threshold gadget in the binary action model.
The result is that the variable gadget ensures
that $S(x_i) \ne S(\lnot x_i)$ and that each consistent
signal configuration gives equal likelihoods of $\theta=\sT$ and $\theta=\sF$.

In the clause gadget (see Figure~\ref{fig:belief-clause}) for a clause
$C_j$
there is an auxiliary agent
observing four agents:
\begin{itemize}
\item Two agents corresponding to literals occurring in $C_j$.
\item Agent $\EVAL$.
\item Agent $E_j$ that receives a private signal
  $S(E_j) \in \{0, 1, 2\}$. For simplicity we relax our model a bit
  and allow a private signal with ternary value. Its signal distribution
  is such that the respective likelihoods satisfy
  \begin{align}
    m_0 &:= \ell_0(E_j) = \ln \frac{\Pr[S(E_j)=0 \mid \Theta=\sT]}
          {\Pr[S(E_j)=0 \mid \Theta=\sF]} \; ,\nonumber\\
    m_1 &:= \ell_1(E_j) = m_0 + D \; ,\nonumber\\
    m_2 &:= \ell_2(E_j) = m_0 + 2D \; , \label{eq:19}
  \end{align}
  where $D := \ell_1-\ell_0 = 2\ln 3$. Furthermore, the probabilities
  $q(\theta_0,b) := \Pr[S(E_j) = b \mid \theta=\theta_0]$
  for $\theta_0 \in \{\sT,\sF\}$
  and $b \in \{0,1,2\}$ are chosen such that
  \begin{align}
    \label{eq:20}
    q(\sT,2)q(\sT,0) = q(\sT,1)^2 = q(\sF,2)q(\sF,0) = q(\sF,1)^2 \; .
  \end{align}
  One checks that \eqref{eq:19} and \eqref{eq:20} are achieved
  (with $m_0 = -D$) by setting
  $q(\sT,2) = q(\sF,0) = q'', q(\sT,1)=q(\sF,1)=q', q(\sT,0)=q(\sF,2)=1-q'-q''$,
  where $(q'', q')$ is the unique positive solution of
  \begin{align*}
    \begin{cases}
      \frac{q''}{1-q'-q''} = 9 \;, \\
      (q')^2 = q'(1-q'-q'') \;,
    \end{cases}
  \end{align*}
  which turns out to be $q'' = 9/13$ and $q' = 3/13$.
\end{itemize}
The auxiliary agent is observed by $\OBS$, broadcasting belief
$A = 3\ell_0 + m_0 + 3D$.

\begin{figure}[ht]\centering
  \caption{Clause gadget.}
  \label{fig:belief-clause}
  \begin{tikzpicture}
    \node [font=\fontsize{15}{0}\selectfont] at (1, 0.2)
    {$C_j = \lnot x_1 \lor x_3$};

    \node [font=\fontsize{6}{0}\selectfont] at (-1.95, -0.6) {$\ell_0, \ell_0+D$};
    \node [draw, circle, label=center:$\lnot x_1$] (x1) at (-2.25, -1.2)
    {\phantom{$\lnot x_n$}};
    \node [font=\fontsize{6}{0}\selectfont] at (0.05, -0.6) {$\ell_0, \ell_0+D$};
    \node [draw, circle, label=center:$x_3$] (x3) at (-0.25, -1.2)
    {\phantom{$\lnot x_n$}};
    \node [font=\fontsize{6}{0}\selectfont, align=center] at (2.05, -0.4)
    {$m_0,m_0+D$\\$m_0+2D$};
    \node [draw, circle, label=center:$E_j$] (Ei) at (1.75, -1.2)
    {\phantom{$\lnot x_n$}};
    \node [font=\fontsize{6}{0}\selectfont] at (4.3, -0.6)
    {$\ell_0, \ell_0+D$};
    \node [draw, ellipse] (EVAL) at (4, -1.2) {$\EVAL$};
    \node [draw, circle] (A) at (1, -3.5) {\phantom{$\lnot x_n$}};
    \node [font=\fontsize{6}{0}\selectfont] at (1, -4.1) {$3\ell_0+m_0+3D$};    
    \node [draw, circle] (OBS) at (1, -6) {$\OBS$};

    \draw [myArrow] (A) [bend left] to (x1);
    \draw [myArrow] (A) [bend left] to (x3);
    \draw [myArrow] (A) [bend right] to (Ei);
    \draw [myArrow] (A) [bend right] to (EVAL);
    \draw [myArrow] (OBS) to (1, -4.3);
  \end{tikzpicture}
\end{figure}

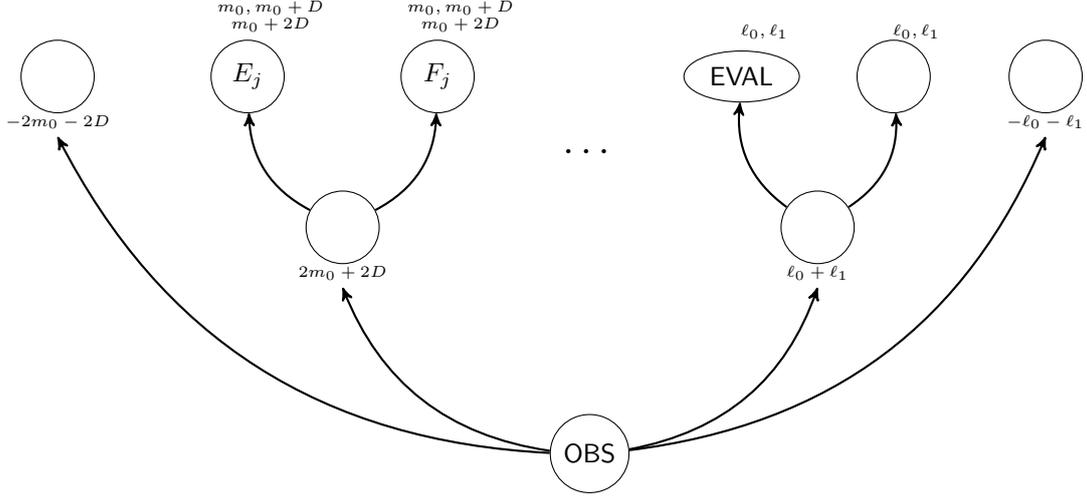
\begin{figure}[ht]\centering
  \caption{Gadgets for $E_j$ and $\EVAL$ agents.}
  \label{fig:ei-eval-gadget}
  \begin{tikzpicture}
    \node [draw, circle] (A) at (-2.5, 10) {\phantom{$\lnot x_n$}};
    \node [font=\fontsize{6}{0}\selectfont] at (-2.5, 9.4) {$-2m_0-2D$};
    \node [font=\fontsize{6}{0}\selectfont, align=center] at (0.3, 10.8)
    {$m_0,m_0+D$\\$m_0+2D$};
    \node [draw, circle, label=center:$E_j$] (Ei) at (0, 10)
    {\phantom{$\lnot x_n$}};
    \node [font=\fontsize{6}{0}\selectfont, align=center] at (2.8, 10.8)
    {$m_0,m_0+D$\\$m_0+2D$};
    \node [draw, circle, label=center:$F_j$] (Ei') at (2.5, 10)
    {\phantom{$\lnot x_n$}};
    \node [draw, circle] (B) at (1.25, 8) {\phantom{$\lnot x_n$}};
    \node [font=\fontsize{6}{0}\selectfont] at (1.25, 7.4) {$2m_0+2D$};

    \node [font=\fontsize{15}{0}\selectfont] at (4.5, 9) {$\cdots$};
    
    \node [font=\fontsize{6}{0}\selectfont] at (6.8, 10.6)
    {$\ell_0, \ell_1$};
    \node [draw, ellipse] (EVAL) at (6.5, 10) {$\EVAL$};
    \node [font=\fontsize{6}{0}\selectfont] at (8.8, 10.6)
    {$\ell_0,\ell_1$};
    \node [draw, circle] (C) at (8.5, 10) {\phantom{$\lnot x_n$}};
    \node [draw, circle] (D) at (10.5, 10) {\phantom{$\lnot x_n$}};
    \node [font=\fontsize{6}{0}\selectfont] at (10.5, 9.4) {$-\ell_0-\ell_1$};
    \node [draw, circle] (E) at (7.5, 8) {\phantom{$\lnot x_n$}};
    \node [font=\fontsize{6}{0}\selectfont] at (7.5, 7.4) {$\ell_0+\ell_1$};

    \node [draw, circle] (OBS) at (4.5, 5) {$\OBS$};

    \draw [myArrow] (B) to [bend left] (Ei);
    \draw [myArrow] (B) to [bend right] (Ei');
    \draw [myArrow] (E) to [bend left] (EVAL);
    \draw [myArrow] (E) to [bend right] (C);
    \draw [myArrow] (OBS) to [bend left] (-2.5, 9.2);
    \draw [myArrow] (OBS) to [bend left] (1.25, 7.2);
    \draw [myArrow] (OBS) to [bend right] (7.5, 7.2);
    \draw [myArrow] (OBS) to [bend right] (10.5, 9.2);
  \end{tikzpicture}
\end{figure}

Since we want to be somewhat more precise in counting likelihoods induced
by different assignments, we introduce additional gadgets ``neutralizing''
likelihoods induced by signals of agents $E_j$ and $\EVAL$,
illustrated in Figure~\ref{fig:ei-eval-gadget}. Their principle is basically
the same as for the variable agents. For example, for each agent $E_j$ we
introduce
another agent $F_j$ with the same signal distribution, an agent observing
both $E_j$ and $F_j$ and broadcasting
$A = 2m_0 + 2D$ to $\OBS$ and yet another agent
broadcasting opposite belief $A = -2m_0 -2D$ to $\OBS$.
In all, these agents ensure that any private signals to $E_j$ and
$\EVAL$ do not affect the likelihood of state of the world $\theta$.

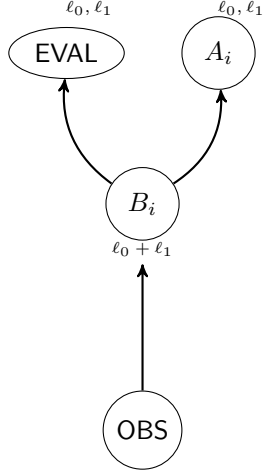
\begin{figure}[ht]\centering
  \caption{One of $K$ parts of the ``amplification'' mechanism.}
  \label{fig:amplification}
  \begin{tikzpicture}
    \node [font=\fontsize{6}{0}\selectfont] at (6.8, 10.6)
    {$\ell_0, \ell_1$};
    \node [draw, ellipse] (EVAL) at (6.5, 10) {$\EVAL$};
    \node [font=\fontsize{6}{0}\selectfont] at (8.8, 10.6)
    {$\ell_0,\ell_1$};
    \node [draw, circle, label=center:$A_i$] (Ai) at (8.5, 10) {\phantom{$\lnot x_n$}};
    \node [draw, circle, label=center:$B_i$] (Bi) at (7.5, 8) {\phantom{$\lnot x_n$}};
    \node [font=\fontsize{6}{0}\selectfont] at (7.5, 7.4) {$\ell_0+\ell_1$};

    \node [draw, circle] (OBS) at (7.5, 5) {$\OBS$};

    \draw [myArrow] (Bi) to [bend left] (EVAL);
    \draw [myArrow] (Bi) to [bend right] (Ai);
    \draw [myArrow] (OBS) to (7.5, 7.2);
  \end{tikzpicture}
\end{figure}

Finally, we let $b := 2N$ and introduce agents
$A_1, \ldots, A_b$ and $B_1, \ldots, B_b$. Each agent $A_i$ receives
a $(p_\sT, p_\sF)$ private signal. Agent $B_i$ observes agents $\EVAL$
and $A_i$ and broadcasts $\ell_0+\ell_1$ to agent $\OBS$
(see~Figure~\ref{fig:amplification}).
This concludes
the description of the reduction. 

\paragraph{Analysis}
The analysis proceeds analogous
to the proof of Theorem~\ref{thm:np}.
First, the network is clearly of size $O(N+M)$ and has the required DAG
structure with significant time $t=2$ for agent $\OBS$.
Next, we convince ourselves
that the private signals consistent with observations
of $\OBS$ can be characterized as:
\begin{itemize}
\item For each assignment $x$ there exists exactly one
  consistent configuration of private signals such that
  $S(\EVAL) = 1$ and $S(A_i) = 0$ for each $i \in \{1,\ldots,b\}$.
\item For each \emph{satisfying} assignment $x$ there is exactly one
  consistent configuration such that $S(\EVAL) = 0$ and $S(A_i) = 1$
  for each $i \in \{1, \ldots, b\}$.
\item There are no other consistent signal configurations.
\end{itemize}

Then, we define $P(x, s, \theta_0)$ as the probability that
$\theta = \theta_0$ and that there arises
the unique signal configuration consistent with assignment $x$
and $S(\EVAL) = s$. The gadgets (recall the relation
\eqref{eq:20} for agents $E_j$ and $F_j$) ensure that $P(\cdot)$
is equal to
\begin{align*}
  P(x, 1, \sT) = q \cdot \left(\frac{1}{4}\right)^b \; , \qquad
  P(x, 1, \sF) = q \cdot \left(\frac{3}{4}\right)^b \; ,
\end{align*}
and, for each assignment $x$ that is satisfying, additionally
\begin{align*}
  P(x, 0, \sT) = q \cdot \left(\frac{3}{4}\right)^b \; , \qquad
  P(x, 0, \sF) = q \cdot \left(\frac{1}{4}\right)^b \; ,
\end{align*}
where $q$ is a universal common factor that depends only on $N$ and $M$.
Recalling that $A$ denotes the number of satisfying assignments
in $\phi$, we can conclude
that the likelihood of agent $\OBS$ at its significant
time $t=2$ is given by
\begin{align*}
  \frac{\mu(\OBS)}{1-\mu(\OBS)}
  &=
  \frac{A\cdot(3/4)^b + 2^N\cdot(1/4)^b}{2^N\cdot(3/4)^b + A\cdot(1/4)^b}
  = \frac{A}{2^N} \cdot \frac{ 1 + \frac{2^N}{A \cdot 3^b}}
    {1 + \frac{A}{2^N\cdot 3^b}}\\
  & \in \frac{A}{2^N} \cdot \left[
    1-\frac{1}{3^b}, 1+\frac{2^N}{3^b}
    \right]
    \subseteq \frac{A}{2^N} \cdot \left[ 1 \pm \frac{1}{4^N} \right] \; .
\end{align*}
In particular,
\begin{align*}
  \left|\frac{\mu(\OBS)}{1-\mu(\OBS)} - \frac{A}{2^N} \right|
  \le \frac{A}{8^N} < \frac{1}{2^{N+1}} \; ,
\end{align*}
so rounding the likelihood to the nearest multiple of $2^{-N}$
successfully recovers the number of satisfying assignments $A$.
\qed

\section{Conclusion}
\label{sec:conclusion}

A natural open question is to make progress on the
approximate-case hardness in one of the models. For example, one could
try to establish
$\NP$-hardness for a worst-case network, but holding for signal configurations
arising with non-negligible probability.
In a different direction, as was mentioned in the introduction, there 
remains a gap between our $\PSPACE$-hardness result and exponential
space required by the best known algorithm.

Another interesting problem arises from trying to extend
our results to the revealed beliefs model, as discussed in
Sections~\ref{sec:belief} and~\ref{sec:sharp}.
Thinking in terms of two-player games, consider a class of
``no-mistakes-allowed'' games: Games where the player with winning strategy
always has exactly one winning move, with all alternative moves in a given
position leading to a losing position (and this property holding recursively
in all positions reachable from the initial one).

Certainly deciding if a position is winning for the first player in such games
is in $\PSPACE$.
On the other hand, since such a game with just the existential player
corresponds to a $\SAT$ formula with zero or one satisfying assignments,
by the Valiant-Vazirani theorem \cite{VV86} it is
also (morally) $\NP$-hard. This leaves a large gap between $\NP$ and $\PSPACE$.  

For example,
suppose we want to prove $\Pi_2$-hardness in the revealed belief model. 
Then it is natural to consider formulas of the form
\begin{align*}
  \forall x \exists y: \phi(x, y) \; ,
\end{align*}
and the question becomes: How hard is it to distinguish between the cases
\begin{itemize}
\item YES: For all $x$, there exists unique $y$ such that $\phi(x, y) = 1$.
\item NO: There exists unique $x_0$ such that no $y$ satisfies $\phi(x_0, \cdot)$.
  For all other $x$, there exists a unique $y$ such that $\phi(x, y) = 1$.
\end{itemize}
How hard is this problem? In particular, can it be shown
to be harder than $\NP$ (in some sense)?
Hardness of such games can be thought of as a generalization
of the Valiant-Vazirani theorem.

\bibliographystyle{alpha}
\bibliography{bibliography}
\end{document}